\documentclass[aps,prb,twocolumn,longbibliography,superscriptaddress,10pt]{revtex4-2} %two column version

\usepackage[utf8]{inputenc}
\usepackage{amssymb,amsmath,amsfonts}
\usepackage{graphicx}
\usepackage{natbib}
\usepackage{multirow}
\usepackage{enumerate}
\usepackage{physics}
\usepackage{gensymb}
\usepackage{subfigure}
\usepackage{hyperref}
\usepackage[dvipsnames]{xcolor}
\usepackage{soul}
\usepackage{comment}
\usepackage[T1]{fontenc}
\usepackage[normalem]{ulem}

\newcommand{\diag}{\text{diag}}
\newcommand{\mi}{\mathrm{i}}
\renewcommand{\vec}[1]{\textbf{\textit{#1}}}

\begin{document}

\title{Classification and magic magnetic-field directions\\ for spin-orbit-coupled double quantum dots}

\author{Aritra Sen}

\affiliation{Department of Theoretical Physics, Institute of Physics, Budapest University of Technology and Economics, M\H{u}egyetem rkp. 3., H-1111 Budapest, Hungary}

\author{Gy\"orgy Frank}

\affiliation{Department of Theoretical Physics, Institute of Physics, Budapest University of Technology and Economics, M\H{u}egyetem rkp. 3., H-1111 Budapest, Hungary}

\author{Baksa Kolok}

\affiliation{Department of Theoretical Physics, Institute of Physics, Budapest University of Technology and Economics, M\H{u}egyetem rkp. 3., H-1111 Budapest, Hungary}

\author{Jeroen Danon}

\affiliation{Department of Physics, Norwegian University of Science and Technology, NO-7491 Trondheim, Norway}

\author{Andr\'as P\'alyi}

\affiliation{Department of Theoretical Physics, Institute of Physics, Budapest University of Technology and Economics, M\H{u}egyetem rkp. 3., H-1111 Budapest, Hungary}

\affiliation{MTA-BME Quantum Dynamics and Correlations Research Group, M\H{u}egyetem rkp. 3., H-1111 Budapest, Hungary}

%\thanks{}
\begin{abstract}
The spin of a single electron confined in a semiconductor quantum dot is a natural qubit candidate. 
Fundamental building blocks of spin-based quantum computing have been demonstrated in double quantum dots with significant spin-orbit coupling. 
Here, we show that spin-orbit-coupled double quantum dots can be categorised in six classes, according to a partitioning of the multidimensional space of their $g$ tensors. 
The class determines physical characteristics of the double dot, i.e., features in transport, spectroscopy, and coherence measurements, as well as qubit control, shuttling, and readout experiments. 
In particular, we predict that the spin physics is highly simplified due to pseudospin conservation, whenever the external magnetic field is pointing to special directions (``magic directions''), where the number of special directions is determined by the class.
We also analyze the existence and relevance of magic loops in the space of magnetic-field directions, corresponding to equal local Zeeman splittings.
These results present an important step toward precise interpretation and efficient design of spin-based quantum computing experiments in materials with strong spin-orbit coupling.
\end{abstract}

\maketitle

\tableofcontents

\section{Introduction}

Double quantum dots (DQDs) are workhorses in the experimental exploration of quantum computing with electron spins~\cite{Loss1998,Hanson2007,Zwanenburg2013,Burkard2021}.
DQDs allowed spin qubit initialization and readout in early experiments based on the Pauli blockade transport effect~\cite{Ono2002,Koppens2006,Nowack2007}.
Since then, numerous experimental demonstrations of single- and two-qubit gates~ \cite{Loss1998,Petta2005,Koppens2006,Nowack2007}, 
qubit readout 
\cite{West2019,Mi2018,Samkharadze2018}, qubit shuttling \cite{Fujita2017,Yoneda2021}, and few-qubit quantum processors \cite{Watson2018,Hendrickx2021,Philips2022} have been completed. 

Spin-orbit interaction often plays a pronounced role in the physical properties of DQDs. 
This is the case, for example, for electrons and holes in III-V semiconductors such as InAs and InSb, or holes in group-IV semiconductors such as Si or Ge. 
Spin-orbit interaction can be an asset or 
a nuisance; for example, it enables coherent electrical spin control \cite{Golovach2006,Nowack2007,Nadjperge2010Nat,Vandenberg2013,Voisin2016,Crippa2018,Froning2021NatNano}, but also contributes to decoherence~\cite{Elzerman2004,Khaetskii2000}.
Hence, understanding spin-orbit-related features and opportunities is of great importance for spin-based quantum computing.

One important consequence of the spin-orbit interaction for spin-based qubits is the renormalization of the $g$ factor.
In fact, an anisotropic Zeeman splitting has been observed in a wide range of experiments~\cite{Voisin2016,NadjPerge2012PRL,Hofmann2017,Scherubl2019,Froning2021,Jirovec2022PRL,Han2023,Piot,HendrickxIBM}, implying that the magnetic response of a spin qubit in a semiconductor with strong spin-orbit coupling must be described by a $g$ tensor, and not by a scalar $g$ factor.
The anisotropy of the $g$ tensors is caused by an interplay~\cite{BulaevPRL2005,Liles,AbadilloUriel} of spin-orbit interaction, electric fields, and mechanical strain, such that these effects are often strongly inhomogeneous due to the symmetry-breaking character of the nanostructured environment of the DQD.
These inhomogeneities, in turn, can include significant nonuniformity of the $g$ tensors in quantum dot arrays, as well as strong misalignment~\cite{Scherubl2019,Geyer2022,Piot} between the principal directions of the $g$ tensor and the high-symmetry crystallographic directions of the sample.

\begin{figure}
\centering
\includegraphics[width=0.5\textwidth]{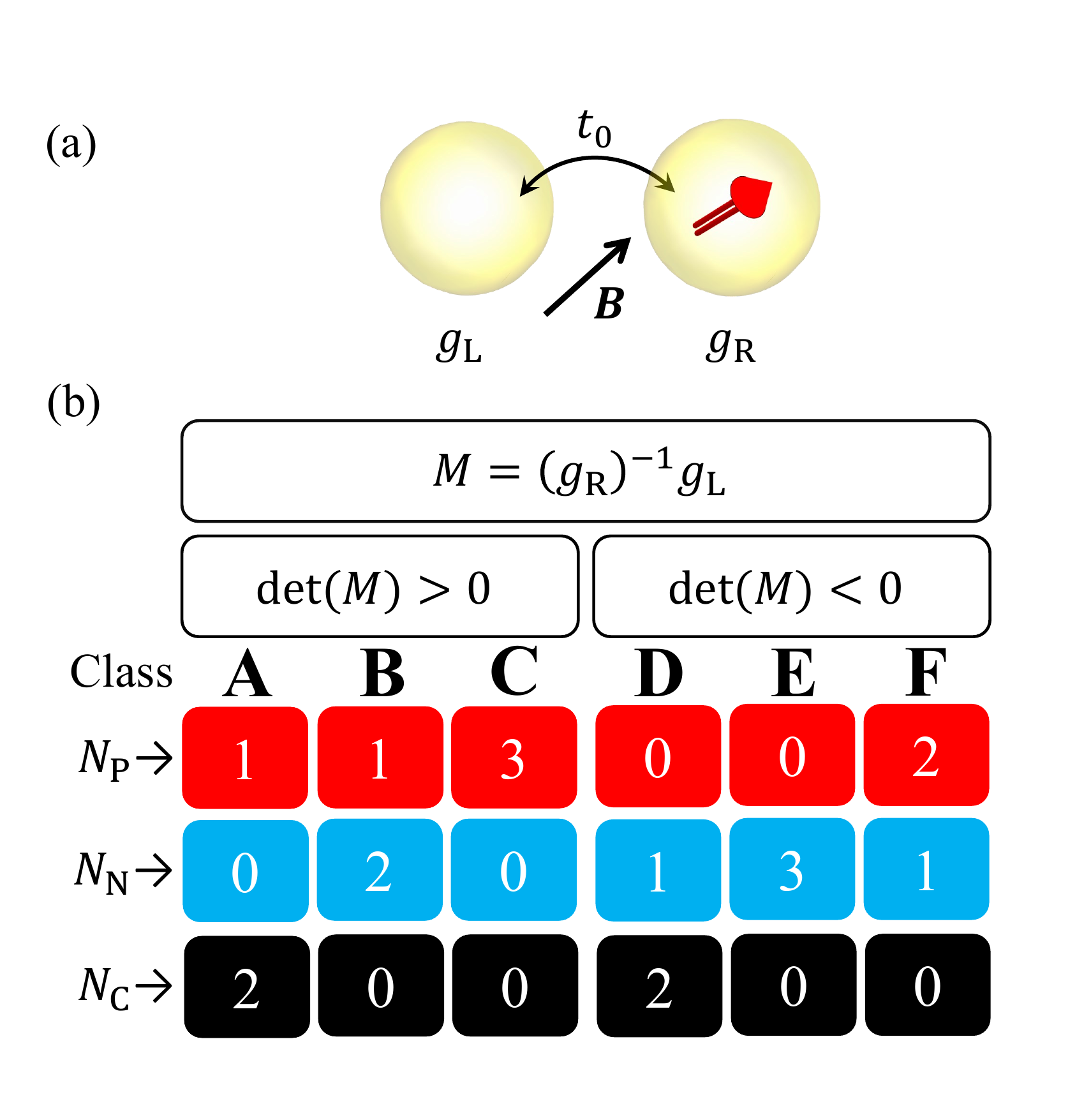}
\caption{Classification of spin-orbit-coupled double quantum dots.
(a) Illustration of a spin-orbit-coupled DQD.
In our preferred gauge, tunneling is pseudospin-conserving ($t_0$), and the $g$ tensors are real (but not necessarily symmetric) matrices. 
(b) Classification based on the eigenvalue structure of the combined $g$ tensor $M$.
Red, blue, and black rows show the number of positive, negative, and complex (nonreal) eigenvalues, respectively, defining the six classes from A to F.
For a given class (column), these numbers add up to three, since $M$ is a $3 \times 3$ matrix.
\label{fig:singleelectronandclasstable}
}
\end{figure}

In this work, we classify spin-orbit-coupled DQDs into six different classes according to 
their \textit{g} tensors, see Fig.~\ref{fig:singleelectronandclasstable}. 
The classification is conveniently carried out in a gauge of `pseudospin-conserving tunneling'. 
In such a gauge, the classification is based on the combined $g$ tensor $M =  g_\textrm{R}^{-1}g_\textrm{L}$ constructed from the $g$ tensors $g_\textrm{L}$ and $g_\textrm{R}$ of the two dots. 
In fact, the classification is defined by the eigenvalue structure of the combined $g$ tensor $M$, i.e., how many of its three eigenvalues are positive, negative, or complex. 

We show that the eigenvectors of $M$ associated with positive or negative eigenvalues specify special,``magic'' magnetic-field directions.
Directing the magnetic field along these magic directions, a conserved pseudospin can be defined, yielding a major simplification of qubit dynamics. 

We highlight pronounced physical features associated with these magic magnetic-field directions: 
(i) spectral crossings in the magnetic-field-dependent and detuning-dependent DQD spectrum, observable via microwave spectroscopy, via pronounced features in quantum capacitance, or via a finite-magnetic-field Kondo effect, 
(ii) prolonged spin relaxation time (relaxation sweet spot), and
(iii) high-fidelity qubit shuttling.
We also discuss related features for two-electron DQDs.

We also discuss the relation between our classification and the often-used theoretical framework where local $g$ tensors are assumed to be isotropic and identical, and spin-dependent tunneling is caused by weak Rashba- and Dresselhaus-type spin-orbit interactions, which together define a single ``spin-orbit field direction''~\cite{Nadjperge2010Nat,Tanttu2019,Pribiag2013,Wang2016UNSW,Hofmann2017,Zarassi2017,Froning2021,Marx2020,Wang2018,Han2023}. 
In fact, one of our classes (Class A) includes that special case, whereas the other five classes we predict and discuss 
correspond to novel, unusual phenomenology as the magnetic-field parameter space is explored.

We also analyze the existence of magic loops in the space of magnetic-field directions, corresponding to equal local Zeeman splittings.
These directions correspond to stopping points in Pauli spin blockade in two-electron DQDs, and provide dephasing sweet spots in single-electron DQDs. 
We prove that the existence of such magic loops depends on the singular values of the matrix $M' = g_\text{L} g_\text{R}^{-1}$ (note the difference with respect to $M$ defined above).

The rest of this paper is structured as follows:
In Sec.~\ref{sec:HamiltonianSOCDQD}, we introduce our parametrized model for the spin-orbit-coupled DQD and transform it for convenience to a specific gauge (`gauge of pseudospin-conserving tunneling').
In Sec.~\ref{sec:classifications}, we provide a classification of our DQD model family, i.e., a partitioning of the parameter space based on the eigenvalue structure of the combined $g$ tensor $M = g^{-1}_\text{R} g_\text{L}$.
In Secs.~\ref{singleelectron} and \ref{twoelectron}, we discuss physical features associated with the magic magnetic-field directions in a DQD with a single electron (with two electrons).
In Sec.~\ref{sec:transitions}, we describe how transitions between the different classes can occur as the $g$ tensors are changed by, e.g., tuning the gate voltages of the DQD.
In Sec.~\ref{sec:magicloops}, we analyze magic loops, i.e., magnetic-field directions where the Zeeman splittings in the two dots are equal.
Finally, we conclude in Sec.~\ref{sec:conclusions}.

\section{Hamiltonian for a spin-orbit-coupled double quantum dot}\label{sec:HamiltonianSOCDQD}

We start with a frequently used phenomenological $4 \times 4$ model Hamiltonian describing a single electron (or hole) in a spin-orbit-coupled DQD.
This Hamiltonian acts on the Hilbert space spanned by the local Kramers basis states in the two dots, $\ket*{L \tilde \Uparrow}$, 
$\ket*{L \tilde \Downarrow}$, 
$\ket*{R \tilde \Uparrow}$, and
$\ket*{R \tilde \Downarrow}$, 
where $L$ and $R$ refer to the two dots, and the arrows refer to local pseudospin basis states that form a local Kramers pair in each dot.

In particular, for the local basis states it holds that they are related by the time reversal operator $\mathcal{T}$, e.g., $\ket{L \tilde{\Downarrow}} = \mathcal{T} \ket{L \tilde{\Uparrow}}$.
For a single quantum dot with spatial symmetries, those spatial symmetries imply a natural choice for the Kramers-pair basis \cite{Chibotaru2008}; however, here we consider DQDs, and assume that all spatial symmetries are broken by the nanostructured environment (e.g., gates, leads), which motivates us to use generic Kramers pairs as described above.

In this basis, our Hamiltonian reads
\begin{subequations}\label{eq:hamiltoniansingleelectron_all}
\begin{align}
    &H = H_\text{os} + H_\text{t} + H_\text{Z},\label{eq:hamiltoniansingleelectron_a}\\
    &H_\text{os} = \frac{\epsilon}{2} \tau_z,\label{eq:hamiltoniansingleelectron_b}\\
    &H_\textrm{t} = \tilde{t}_0 \tau_x + \tilde{\vec{t}} \cdot \tilde{\boldsymbol{\sigma}} \otimes \tau_y,\label{eq:hamiltoniansingleelectron_c}\\
    &H_\textrm{Z} = \frac{1}{2} \mu_B \left(\tilde{\boldsymbol{\sigma}}_\textrm{L} \cdot \tilde{g}_\textrm{L} \vec{B}  + \tilde{\boldsymbol{\sigma}}_\textrm{R} \cdot \tilde{g}_\textrm{R} \vec{B}\right), \label{eq:hamiltoniansingleelectron_d}
\end{align}
\end{subequations}
where $H_\text{os}$, $H_\textrm{t}$, $H_\text{Z}$ are on-site, tunneling and Zeeman terms, respectively. 

The vector $\tilde{\boldsymbol{\sigma}}=(\tilde{\sigma}_x,\tilde{\sigma}_y,\tilde{\sigma}_z)$ is the vector of Pauli matrices acting on the local Kramers bases on the two dots, e.g., $\tilde{\sigma}_z = \dyad*{\tilde{\Uparrow}} - \dyad*{\tilde{\Downarrow}}$. 
The vector $\boldsymbol{\tau} = (\tau_x,\tau_y,\tau_z)$ is the vector of Pauli matrices acting on the orbital degree of freedom, e.g., $\tau_z = \dyad*{L} - \dyad*{R}$. 
The vector $\tilde{\boldsymbol{\sigma}}_\textrm{L}$ consists of components such as $\tilde{\sigma}_z \otimes (1+\tau_z)/2$, etc. 
Furthermore, $\epsilon$ denotes the on-site energy detuning between the two dots. The pseudospin-conserving tunneling amplitude is denoted by $\tilde{t}_0$, and $\tilde{\vec{t}}=(\tilde{t}_x,\tilde{t}_y,\tilde{t}_z)$ is the vector of pseudospin-nonconserving tunneling amplitudes. 
Note that the tunneling amplitudes are gauge dependent, such that their combination $\tilde{t}_0^2 + \tilde{\vec{t}}^2$ is gauge invariant.
This tunneling Hamiltonian $H_\textrm{t}$ respects time-reversal symmetry \cite{Danon2009}, the latter being represented by $i \tilde{\sigma}_y K$, where $K$ is complex conjugation.

In the Zeeman term $H_\textrm{Z}$, the Bohr magneton is denoted as $\mu_B$, whereas $\tilde{g}_\textrm{L}$ and $\tilde{g}_\textrm{R}$ are the $g$-tensors of the two dots, and $\vec{B}$ is the external magnetic field.
Note that the matrix elements of the $g$-tensors depend on the gauge choice, i.e., the choice of the local Kramers-pair basis, which we have not yet specified~\cite{Chibotaru2008}.
In a generic gauge, the $g$-tensors are real matrices, but they are not necessarily symmetric.

For convenience, we convert the Hamiltonian above to a gauge that we refer to as the `gauge of pseudospin-conserving tunneling'. 
This is done by a local change of the Kramers basis in one of the dots, say, the right one, i.e., 
$\ket*{R \Uparrow} = W \ket*{R \tilde\Uparrow}$ and 
$\ket*{R \Downarrow} = W \ket*{R \tilde\Downarrow}$, where $W$ is a $2\times 2$ special unitary matrix. 
An appropriately chosen basis change $W$ yields (see Appendix~\ref{appGaugeTransformation} for details) the following transformed Hamiltonian:
\begin{equation}\label{eq:hamiltoniangaugesingleelectron}
H = \frac{\epsilon}{2} \tau_z + t_0 \tau_x + 
\frac{1}{2} \mu_B \left(\boldsymbol{\sigma}_\textrm{L} \cdot g_\textrm{L} \vec{B}  + \boldsymbol{\sigma}_\textrm{R} \cdot g_\textrm{R} \vec{B}\right).
\end{equation}
As a result of the basis change on dot $R$, the corresponding $g$ tensor has been rotated such that $g_\textrm{R} = R \tilde{g}_\textrm{R}$. 
On the other hand, the $g$ tensor of dot $L$ is unchanged, $g_\textrm{L} = \tilde{g}_\textrm{L}$.
The Hamiltonian in Eq.~\eqref{eq:hamiltoniangaugesingleelectron} is illustrated in Fig.~\ref{fig:singleelectronandclasstable}.
In what follows, we refer to 
$g_\textrm{L} \vec B$
and 
$g_\textrm{R} \vec B$
as the \emph{internal Zeeman fields}.
We emphasize that, in this gauge, all effects of spin-orbit interaction are incorporated in the two effective $g$-tensors, and the interdot tunneling term is pseudospin-conserving.
Note that $W$ being an SU(2) matrix implies that in the new gauge the form of the time-reversal operator is preserved, i.e., it is $i\sigma_y K$.

Before analyzing Hamiltonian \eqref{eq:hamiltoniangaugesingleelectron}, let us discuss a few experimental observations regarding $g$ tensors in DQDs.
Based on, e.g., Refs.~\onlinecite{NadjPerge2012PRL,Crippa2018,Scherubl2019,Jirovec2022PRL, HendrickxIBM}, $g$-tensor principal values in semiconductor DQDs range between 0.05 to 30, and the principal axis might~\cite{Jirovec2022PRL,HendrickxIBM} or might not~\cite{Scherubl2019} be correlated with the device geometry. 
Furthermore, in Ref.~\onlinecite{Jirovec2022PRL}, a planar Ge hole DQD was studied, with the conclusion that the $g$ factors in the out-of-plane direction have the same sign on the two dots, whereas they exhibit opposite signs in a certain in-plane direction. 
This anticipates that in DQDs with strong spin-orbit interaction, $g$-tensors can have a rich variety, including strong anisotropy, large $g$-tensor difference between the two dots, and even different signs of the two $g$-tensor determinants are possible. 
From now on, we take these features as our motivating starting point, and discuss potential scenarios arising from this rich variety of $g$-tensor configurations on a conceptual level. 
In this work, we suppress further material-specific considerations, e.g., based on real-space models of strong spin-orbit interaction. 
Such considerations are important steps to be taken in future work.

\section{Magic magnetic field directions and the classification of the combined \MakeLowercase{\textit{g}}-tensor}\label{sec:classifications}

Equation \eqref{eq:hamiltoniangaugesingleelectron} describes a Hamiltonian family parametrized by 20 parameters, out of which 18 describe the two $g$ tensors.
We now classify this Hamiltonian family into six classes. 
The classification is based on the two $g$-tensors. 
In particular, it is based on the physical intuition that there might be special (``magic'') magnetic-field directions such that the internal Zeeman fields $g_\textrm{L} \vec{B}$ and $g_\textrm{R} \vec{B}$ in the two dots are parallel.
If the magnetic field is pointing to such a magic direction, then the pseudospin (more precisely, its projection on the internal Zeeman field direction) is conserved, leading to a major simplification of the spectral and dynamical properties, as discussed below.

For which magnetic-field directions are the internal Zeeman fields $g_\textrm{L} \vec B$ and $g_\textrm{R} \vec B$ parallel to each other?
They are parallel \cite{Frank2020}, i.e., 
\begin{equation}
g_\textrm{L} \vec B \parallel g_\textrm{R} \vec B,
\end{equation}
if it holds that 
\begin{equation}
g_\textrm{R}^{-1} g_\textrm{L} \vec B \parallel \vec B. 
\end{equation}
This holds if $\vec B$ is a (right) eigenvector of the combined $g$ tensor
\begin{equation}\label{eq:Mdef}
    M = g_\textrm{R}^{-1} g_\textrm{L},
\end{equation}
that is, 
\begin{equation}
M \vec{B} = \lambda \vec B.
\end{equation}
In fact, the internal Zeeman fields are aligned (anti-aligned), if $\vec{B}$ is an eigenvector of $M$ corresponding to a positive (negative) eigenvalue. We call the eigenvectors of $M$ corresponding to real eigenvalues as \emph{magic magnetic field directions}.

\begin{figure}
\includegraphics[width=0.48\textwidth]{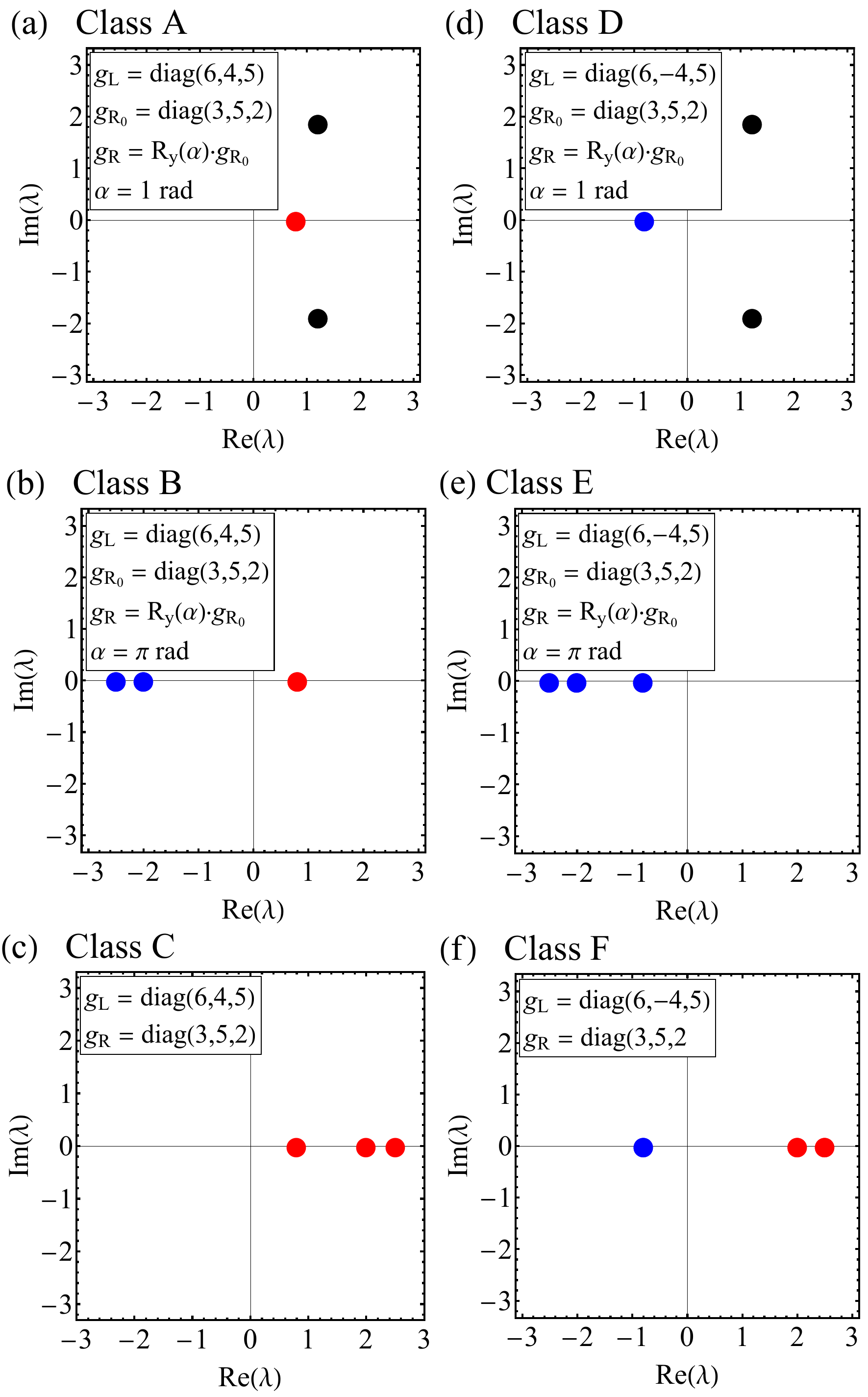}
\caption{Eigenvalues of the combined $g$-tensor $M = g_\textrm{R}^{-1} g_\textrm{L}$, illustrating the sixfold classification of spin-orbit-coupled double quantum dots.
Insets specify the $g$-tensor examples. 
Red, blue, and black points denote positive, negative, and complex eigenvalues, respectively.
For classes A, B, D, and E, $g_R$ is defined via a rotation matrix $R_y(\alpha)$ of angle $\alpha$ around the $y$ axis. 
Note that for classes A and D, the example $g_R$ is nonsymmetric.}
\label{fig:eigenexample}
\end{figure}

The above observation implies that the spin-orbit-coupled DQDs characterized by the Hamiltonian of Eq.~\eqref{eq:hamiltoniangaugesingleelectron} can be categorized into six classes, as shown in Fig.~\ref{fig:singleelectronandclasstable}(b). 
(i) If $\text{det} M >0$, i.e., if the determinants of the two $g$ tensors have the same sign, then there are three classes, to be denoted by A $(+,c,c)$, B $(+,-,-)$, and C $(+,+,+)$.
Here, $+$ stands for a positive eigenvalue, $-$ stands for a negative eigenvalue, and $c$ stands for a complex (nonreal) eigenvalue of the $M$ matrix. 
(ii) If $\text{det} M < 0$, i.e., if the determinants of the two \textit{g}-tensors have opposite signs, then there are three further classes: D $(-,c,c)$, E $(-,-,-)$, and F $(+,+,-)$.
We illustrate these classes in Fig.~\ref{fig:eigenexample} by plotting the eigenvalues of $M$, computed for representative $g$-tensor examples.

Our conclusion so far is that spin-orbit-coupled DQDs can be classified through the eigenvalue characteristics of the combined $g$ tensor $M$. 
The number of positive, negative and complex eigenvalues of $M$ varies as we move between the classes. 
For each real eigenvalue of $M$, there is a magic magnetic-field direction, specified by the corresponding eigenvector of $M$, where pseudospin is conserved. 
Below, we show that the DQD's physical properties depend markedly on the sign of the eigenvalue corresponding to the magic direction, i.e., the case of aligned internal Zeeman fields is accompanied by different physical consequences than the case of anti-aligned internal Zeeman fields.

We also note that in the above classification, we implicitly assumed that the $g$ tensors are invertible, i.e., all eigenvalues are nonzero. 
Nevertheless, our classification is satisfactory in the sense that $g$ tensors with a zero eigenvalue form a zero-measure set within the space of $g$ tensors.
Also, our classification has a certain `robustness' or `stability': given that a Hamiltonian is in a certain class, then its perturbation cannot change the class as long as the perturbation is sufficiently weak. 
Transitions between different classes upon continuous perturbations are discussed in Sec.~\ref{sec:transitions} below.

We highlight already here that certain physical features of a spin-orbit-coupled DQD are determined by an alternative combination of the $g$ tensors, i.e., by the matrix $M' = g_\text{L} g_\text{R}^{-1}$, which we introduce and analyze in Sec.~\ref{sec:magicloops}. 
The eigenvalues of $M'$ are the same as those of $M$, i.e., knowing $M'$ also implies the class of a DQD.
However, the implications we can draw from $M$ and $M'$ are different, as we discuss in more detail in Sec.~\ref{sec:magicloops}.

Finally, we relate our classification to the concept of ``spin-orbit direction'' (or `spin-orbit field direction') \cite{Nadjperge2010Nat,Tanttu2019,Pribiag2013,Wang2016UNSW,Hofmann2017,Zarassi2017,Froning2021,Marx2020,Wang2018,Han2023}.
We note that this phrase may have multiple interpretations;
here we focus on a specific theoretical interpretation.
The model family we study incorporates the special case of weak spin-orbit interaction induced by band-structure effects, e.g., of Rashba or Dresselhaus type. 
In that case, the $g$ tensors of our DQD Hamiltonian are well approximated as identical and isotropic $g$ tensors, and the direction of the tunneling vector $\vec{t}$, often called the `spin-orbit field direction' in this case, is governed by the Rashba and Dresselhaus Hamiltonians.
This special case is included in class A of our classification.
Furthermore, in this case the spin-orbit field direction coincides with the single magic magnetic-field direction.
However, strong spin-orbit interaction and the symmetry breaking associated with the nanostructured environment often implies highly anisotropic $g$ tensors, opening up a broader spectrum of physical scenarios, as characterized by the further five classes of our sixfold classification.

\section{Single-electron effects with magic magnetic-field directions}\label{singleelectron}

In what follows, we highlight the role of the magic magnetic-field directions in determining physical properties.
In this section, we focus on the properties of spin-orbit-coupled DQDs hosting a single electron.

\begin{figure}
\centering
\includegraphics[width=0.48\textwidth]{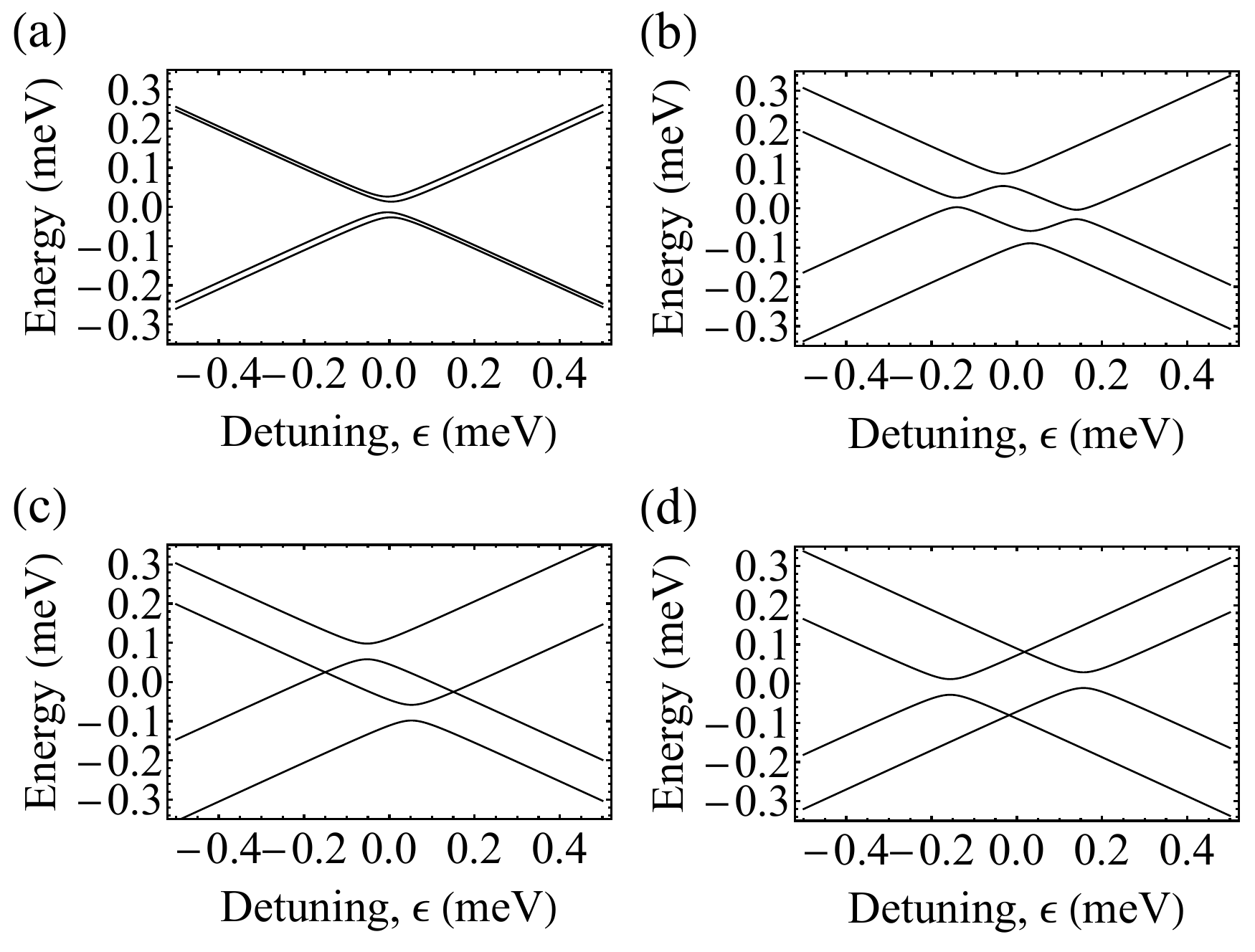}
\caption{Single-electron spectral degeneracies in a spin-orbit-coupled double quantum dot when the magnetic field is in a magic direction. The values of the \textit{g}-tensors in left and right dots are $g_\textrm{L}=\diag(6,-4,5)$ and $g_\textrm{R}=\diag(3,5,2)$, respectively. 
The tunneling amplitude is set to $t_0=0.02$~meV. 
(a) $\vec{B} = (0.05,0,0)$~T, implying that the Zeeman energy
is much smaller than tunneling energy. Note that this field points to a magic direction with a positive eigenvalue.
This yields no spectral degeneracies.
In panels (b) and (c), the magnetic field is  $B = 0.6$~T, implying that the Zeeman energy
is larger than tunneling energy. 
(b) In a generic magnetic-field direction $\vec B/B = (1/\sqrt 2, 1/\sqrt 2, 0)$,
four anticrossings appear; there are no band crossing points.
(c) $\vec B = (0.6,0,0)$~T, i.e., 
$\vec B$ is an eigenvector of $M$ with a positive eigenvalue. 
Band crossings occur between the antibonding pseudospin band and the bonding pseudospin band.
(d) $\vec{B} = (0,0.6,0)$~T, 
i.e., $\vec B$ is an eigenvector of $M$ with a negative eigenvalue. 
A band crossing occurs between the two bonding bands, and another band crossing between the two antibonding bands. Such crossings occur also if the Zeeman splitting is much smaller than the tunneling energy.
}
\label{fig:alignedantialigned}
\end{figure} 

\subsection{Robust spectral degeneracies}
\label{sec:singleelectrondegeneracies}

First, we describe single-electron spectral degeneracies that appear when the magnetic field points to a magic direction. 
These are illustrated in Fig.~\ref{fig:alignedantialigned}, where we show four energy spectra, plotted as a function of the on-site energy detuning $\epsilon$.

% i.e., all crossings seen for the magic directions are changed to anticrossings. 

Figure \ref{fig:alignedantialigned}(a) shows four energy levels (bands) as a function of detuning, 
for a weak magnetic field, such that the Zeeman energy is much smaller than the tunneling amplitude, and the field points into a magic direction with a positive eigenvalue.
In this case, there are no degeneracies in the spectrum. 
Note that the tunneling amplitude sets the gap between the bonding (lower-energy) and antibonding (higher-energy) bands at zero magnetic field.
% There are no degeneracies in Fig.~\ref{fig:alignedantialigned}(a).

By varying the magnetic field strength and direction, we see band anticrossings or crossings, as shown in 
Figs.~\ref{fig:alignedantialigned}(b)--\ref{fig:alignedantialigned}(d).
In Fig.~\ref{fig:alignedantialigned}(b) we show the detuning-dependent spectrum for a generic (nonmagic) magnetic field direction, where the Zeeman splitting is comparable to the tunneling energy. In this case, the bands exhibit four anticrossings, i.e., there are no spectral degeneracies.

Spectral degeneracies are associated with the magic magnetic-field directions, as shown in Figs.~\ref{fig:alignedantialigned}(c)--\ref{fig:alignedantialigned}(d).
In Fig.~\ref{fig:alignedantialigned}(c), the magnetic field points to a magic direction of a positive eigenvalue.
In this case, two band crossing points between the bonding band of the high-energy pseudospin and the antibonding band of the low-energy pseudospin are present.
The reason for the presence of these spectral crossing points is that the $4\times 4$ Hamiltonian
now separates into two uncoupled pseudospin sectors, due to pseudospin conservation, which in turn is the consequence of the magnetic field pointing to a magic direction. 

In Fig.~\ref{fig:alignedantialigned}(d), the magnetic field points to a magic direction of a negative eigenvalue.
In this case, there is a crossing point between the two bonding bands, and there is a crossing point between the two antibonding bands. 
Again, the crossings arise due to pseudospin conservation. 
We emphasize that the sign of the eigenvalue (corresponding to the magic direction along which the magnetic field is applied) determines which pair(s) of bands cross. 

Remarkably, these spectral degeneracies are robust in the following sense:
If the $g$ tensors suffer a small perturbation, e.g., due to a small change of the voltages of the confinement gates, then the eigenvalue characteristics (number of positive, negative, complex eigenvalues) of the combined $g$ tensor $M$ remain unchanged, albeit that the eigenvectors and eigenvalues of $M$ do suffer a small change. 
This means that the magic directions change a bit, but a small adjustment of the magnetic field to align with the new magic direction is sufficient to reinstate the degeneracy points in the detuning-dependent spectrum again. 

This robustness of the degeneracy points is often phrased as topological protection \cite{Armitage2018,Riwar2016,Scherubl2019,Stenger2019},
and it is a direct consequence of the fact that the subset (``stratum'') of matrices with a twofold eigenvalue degeneracy has a codimension of three in the space of Hermitian matrices \cite{vonNeumann1929,arnold1995}. 

In fact, we can consider a three-dimensional parameter space formed by the detuning $\epsilon$, and the polar and azimuthal angles $\theta$ and $\phi$ that characterize the direction of the magnetic field. 
In that three-dimensional parameter space, one can associate a topological invariant to the degeneracy point, which is often called the Chern number \cite{Asboth2016} 
or the local degree \cite{pinter2022}.
For Hermitian matrices parametrized by three parameters (such as our Hamiltonian), (i) band crossings arise generically, (ii) the value of the Chern number associated with a generic band crossing is $\pm 1$, and (iii) such band crossings are robust against small changes of further control parameters (e.g., the elements of the $g$-tensors, or the magnetic field strength, in our physical setup).
We have computed the Chern number for the band crossings shown in Fig.~\ref{fig:alignedantialigned}, and indeed found $\pm 1$, confirming the topological protection of these degeneracy points. 

A natural question is how to perform the classification experimentally?
In other words, given an experimental setup with a tuned-up single-electron DQD in a material with strong spin-orbit coupling, how could an experiment find out the eigenvalue class corresponding to that setup?

(1) A natural idea is to use spectroscopy based on electron-spin resonance\cite{Koppens2006} or electrically driven spin resonance\cite{Nowack2007}, such that the magnetic field strength and direction are scanned.
In principle, these techniques provide access to all spectral gaps as functions of detuning and magnetic field, and hence are suited to locate the spectral degeneracies in the parameter space, e.g., in the space of $\epsilon$, $\theta$, and $\phi$.
On the one hand, the number of degeneracy points found between the lowest two energy levels is equal to the number of degeneracy points between the highest two energy levels, and this number is also the number of negative eigenvalues of $M$.
On the other hand, the number of degeneracy points between the first and second excited levels implies the number of positive eigenvalues of $M$, hence completing the experimental classification.

(2) Besides resonant mapping of the energy gaps via the spectroscopic techniques described above, the magic directions belonging to the negative eigenvalues of the combined $g$-tensor can also be found using simpler techniques sensitive to the ground state only. 
A ground-state degeneracy point, such as the one depicted in Fig.~\ref{fig:alignedantialigned}(c), is often detected via cotunneling spectroscopy \cite{Scherubl2019}. 
Moreover, at low enough temperature this degeneracy causes a Kondo effect at finite magnetic field  \cite{Nygaard2016Kondo,Scherubl2019}. 
Finally, the ground-state degeneracy point of  Fig.~\ref{fig:alignedantialigned}(c) leads to characteristic features of the quantum capacitance, e.g., the suppression of the latter \cite{Han2023} compared with the quantum capacitance induced by an anticrossing.
This quantum capacitance suppression can be detected as a function of the magnetic field direction, along the lines of the experiment of Ref.~\cite{Han2023}, revealing the magic direction belonging to the negative eigenvalue. 
We discuss this effect further in Appendix~\ref{app:capacitance}.

Methods (1) and (2) can help to do the classification experimentally.
However, it can happen that the measured data suggest a level crossing, even when there is no level crossing but an anticrossing, with a local minimum of the spectral gap that is nonzero and smaller than the spectral resolution of the experiment.
Such cases might hinder a successful classification.

What are the practical aspects of this caveat?
Below, we describe certain functionalities in spin-qubit experiments that we associate to magic magnetic-field directions.
Some of these functionalities rely only on the vanishing (or almost-vanishing) anticrossing size between neighboring levels: an example is the fully diabatic transition utilized for spin-qubit readout, see Sec.~\ref{sec:psb}.
For some other functionalities, such as the relaxation sweet spot of Sec.~\ref{sec:relaxationsweetspot} and the shuttling sweet spot in Sec.~\ref{sec:shuttlingsweetspot}, the almost-vanishing anticrossing size is not sufficient; for these, the magnetic field has to point into a magic direction.
In fact, these sweet-spot functionalities serve as further evidence for a magic direction: if observed, they establish confidence that the apparent level crossing is an actual level crossing belonging to a magic direction, even if the methods of (1) and (2) have insufficient spectral resolution.

\subsection{Relaxation sweet spot}
\label{sec:relaxationsweetspot}

A further physical consequence of setting the magnetic field in a magic direction is an increased spin-relaxation time.
That is, the magic direction provides a \emph{spin-relaxation sweet spot} in the parameter space of magnetic-field directions.
The description of this feature is as follows.

In a spin-orbit-coupled DQD, a key mechanism of spin relaxation is detuning noise. 
Electric fluctuations, including phonons, fluctuating charge traps, gate voltage jitter, etc., induce on-site energy fluctuations, leading to fluctuations of the detuning $\epsilon$. 
In turn, these detuning fluctuations push the electron back-and-forth between the two dots. 
If the magnetic field is not along a magic direction, then electron feels an internal Zeeman field with a fluctuating direction, leading to qubit relaxation. 
However, if the magnetic field is pointing along a magic direction, then the pseudospin is conserved despite the fluctuating electron motion, and hence qubit relaxation is suppressed.

Of course, relaxation is absent only in the idealized case described above. 
In reality, electric fluctuations not only modify the detuning, but also reshape the landscape of the double-dot confinement potential, and hence modify tunneling as well as the $g$-tensors. 
Nevertheless, as long as the dominant qubit relaxation mechanism is due to detuning noise, a qubit relaxation sweet spot is expected if the magnetic field is pointing in a magic direction.

\subsection{Shuttling sweet spot}
\label{sec:shuttlingsweetspot}

Shuttling electrons in quantum dot arrays is a prominent element of proposals describing scalable spin qubit architectures \cite{RoyLi2018Science,Helsen2018iop,QMhamza2022,QMcrawford2022axv,QMpatomaki2023}. 
In such architectures, it is desirable to preserve the quantum state of a spin qubit upon shuttling to a neighboring dot \cite{Yoneda2021,Ginzel2020,Gullans2020prb,Langrock2023prx,Krzywda2020noise,Krzywda2021prb,Mortemousque}.
In a double-dot setup, such high-fidelity qubit shuttling is facilitated if a conserved pseudospin can be defined. 
This is indeed the case, whenever the magnetic field is oriented along a magic direction.

\section{Two-electron effects with magic magnetic-field directions}\label{twoelectron}

So far we studied a spin-orbit-coupled DQD hosting a single electron, and we investigated the role of the magic magnetic-field directions in the qualitative structure of the single-particle spectrum, as well as their relation to sweet spots for relaxation and coherent shuttling.
However, such DQD systems are also often operated in the \emph{two}-electron regime, typically tuned to the vicinity of the $(1,1)$--$(0,2)$ charge degeneracy.

Measurement of the current through the DQD in this setting is useful to characterize both coherent and dissipative components of the spin dynamics.
A combination of DQD gate-voltage pulse sequences and charge sensing provides elementary experiments toward spin-based quantum information processing, demonstrating initialization, coherent control, readout, and rudimentary quantum algorithms.

The mechanism of Pauli spin blockade (PSB) is an essential ingredient in those experiments. 
In this section, we connect the two-electron DQD physics and PSB to the matrix $M$ defined in Sec.~\ref{sec:classifications}.
We assess the potential of the spin-orbit-coupled DQD for hosting spin qubits and performing PSB-based qubit readout, highlighting the special role of the magic magnetic-field directions we introduced above.
We believe that connecting the experimental phenomenology to the properties of the matrix $M$ provides a more precise representation of the underlying physics than the usual interpretation in terms of a spin-orbit field that only acts during electron tunneling.
In particular, we provide a potential explanation of the recently observed experimental feature \cite{Hendrickx2021} which we term ``inverted PSB readout''.

\subsection{Robust spectral degeneracies in the two-electron low-energy spectrum}

First, we investigate the low-energy part of the spectrum close to the $(1,1)$--$(0,2)$ charge transition.
To this end, we write a two-electron version of the Hamiltonian \eqref{eq:hamiltoniangaugesingleelectron}, projected to the four $(1,1)$ states---$\ket{{\cal T}_+} = \ket{L\Uparrow,R\Uparrow}$, $\ket{{\cal T}_0} = \frac{1}{\sqrt 2} \big[ \ket{L\Uparrow,R\Downarrow} + \ket{L\Downarrow,R\Uparrow} \big]$, $\ket{{\cal T}_-} = \ket{L\Downarrow,R\Downarrow}$, and $\ket{{\cal S}} = \frac{1}{\sqrt 2} \big[ \ket{L\Uparrow,R\Downarrow} - \ket{L\Downarrow,R\Uparrow} \big]$---and the $(0,2)$ singlet $\ket{{\cal S}_{02}} = \frac{1}{\sqrt 2} \big[ \ket{R\Uparrow,R\Downarrow} - \ket{R\Downarrow,R\Uparrow} \big]$, yielding
\begin{align}
\label{eq:twoelectronhamiltonian}
    H^{(2)} = {} & {} \frac{1}{2} \mu_B \left(\boldsymbol{\sigma}_\textrm{L} \cdot g_\textrm{L} \vec{B}  + \boldsymbol{\sigma}_\textrm{R} \cdot g_\textrm{R} \vec{B}\right) \nonumber\\
    {} & {} + \sqrt 2 t_0 \big[ \ket{{\cal S}}\bra{{\cal S}_{02}} + {\rm H.c.} \big] - \epsilon \ket{{\cal S}_{02}}\bra{{\cal S}_{02}}.
\end{align}

\begin{figure}[t]
\centering
\includegraphics[width=0.48\textwidth]{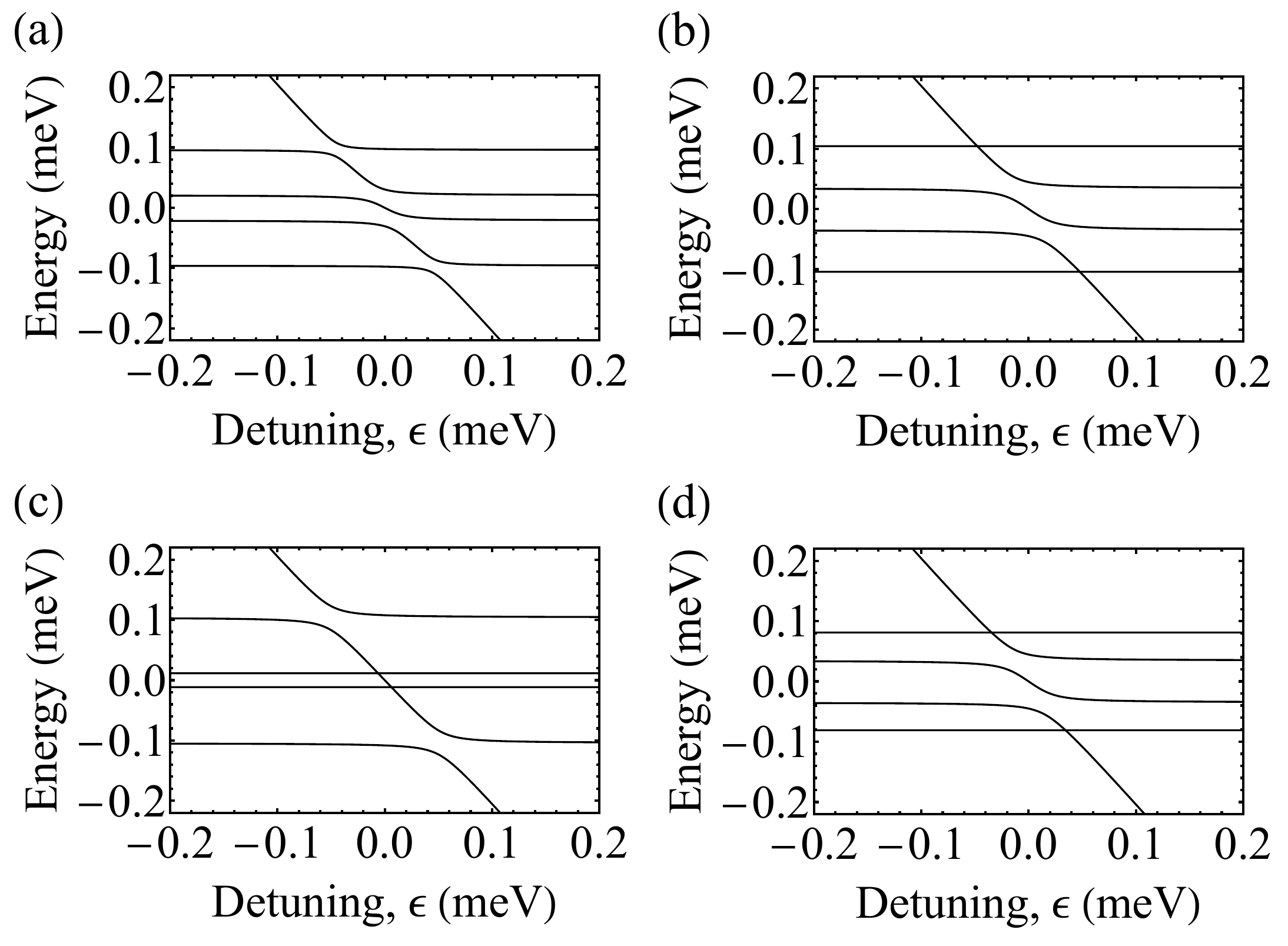}
\caption{Two-electron spectrum of $H^{(2)}$ as a function of detuning $\epsilon$, for a pair of $g$-tensors in class F.
The explicit $g$-tensors used for the left and right dot are, respectively, $g_\textrm{L} = \diag (6, -4, 5)$ and $g_\textrm{R} = \diag (3, 5, 2)$.
The tunneling amplitude is set to ${t}_0 = 0.02$ meV and we use a magnitude $B = 0.2$~T for the applied magnetic field throughout.
(a) The spectrum for a generic direction of $\vec{B} = (\frac{1}{\sqrt 2}, \frac{1}{\sqrt 2}, 0)B$.
In this scenario $\ket{{\cal S}_{02}}$ anticrosses with all four 
$(1,1)$ states.
(b), (d) The magnetic field $\vec{B}$ is oriented along the $x$ and $z$ axes, respectively.
In these two magic directions, the internal Zeeman fields in the two dots are aligned, resulting in two level crossings in the spectrum, between $\ket{{\cal S}_{02}}$ and the highest- and lowest-energy $(1,1)$ states.
(c) Now $\vec{B}$ is oriented along the $y$-axis, resulting in two internal Zeeman fields that are anti-aligned, and we see level crossings between $\ket{{\cal S}_{02}}$ and the two inner $(1,1)$ states.
}
\label{fig:twoelectronspectraclassf}
\end{figure}

To understand the different possible scenarios, we focus on class F, which hosts both types of magic field directions (both corresponding to positive and negative eigenvalues of $M$).
Figure~\ref{fig:twoelectronspectraclassf}(a) shows a typical spectrum as a function of the detuning $\epsilon$, at a finite magnetic field in a generic direction (see the caption for parameters used).
The detuning-dependent state $\ket{{\cal S}_{02}}$ decreases in energy with increasing $\epsilon$ and it anticrosses with all four $(1,1)$ states, indicating that they indeed all have a finite singlet component.
Away from the anticrossings, the four $(1,1)$ eigenstates correspond to the four possible configurations with both pseudospins aligned or anti-aligned with the local internal Zeeman field $g_{\rm L,R}\vec{B}$.

In Figs.~\ref{fig:twoelectronspectraclassf}(b)--\ref{fig:twoelectronspectraclassf}(d) we explore the three magic magnetic-field directions available in class F.
For the simple example $g$ tensors we chose [$g_\textrm{L} = \diag (6, -4, 5)$ and $g_\textrm{R} = \diag (3, 5, 2)$] the three magic directions are along the three Cartesian axes $\hat x$, $\hat y$, and $\hat z$.
Directing $\vec{B}$ along $\hat x$ or $\hat z$ [shown in Fig.~\ref{fig:twoelectronspectraclassf}(b) and \ref{fig:twoelectronspectraclassf}(d), respectively] corresponds to a positive eigenvalue of $M$, causing the local (internal) Zeeman fields on the dots $g_{\rm L,R}\vec{B}$ to be parallel.
This implies that the highest (lowest) $(1,1)$ state in the spectrum has its two pseudospins
parallel to each other, which explains why they do not hybridize with the singlet $\ket{{\cal S}_{02}}$.
In Fig.~\ref{fig:twoelectronspectraclassf}(c) $\vec{B}$ points along $\hat y$, which is a magic direction corresponding to a negative eigenvalue of $M$.
In this case the internal Zeeman fields $g_{\rm L,R}\vec{B}$ are anti-aligned and the highest and lowest $(1,1)$ states now have their pseudospins anti-aligned with each other [each pseudospin aligns with the local internal Zeeman field].
These two states now have a finite overlap with $\ket{\cal S}$ and thus hybridize with $\ket{{\cal S}_{02}}$, whereas the two central states now have parallel pseudospins and thus cross with $\ket{{\cal S}_{02}}$.

We note that the spectral crossings discussed here are robust in the same sense as described for the single-electron spectral crossings in Sec.~\ref{sec:singleelectrondegeneracies}.

\subsection{Single-spin readout via Pauli spin blockade}
\label{sec:psb}

A DQD, hosting two electrons, tuned to the vicinity of the (1, 1)--(0, 2) charge transition, can be used to perform readout of a single-spin qubit via spin-to-charge conversion. 
This readout functionality relies on the PSB mechanism, as we summarize below.

Assume that deep in the (1, 1) charge configuration (left side of the plots in Fig.~\ref{fig:twoelectronspectraclassf}) the left electron is in an unknown pseudospin state, which one wants to read out in the basis of the local pseudospin eigenstates $\ket{+L}$ and $\ket{-L}$, where $+$($-$) denotes the pseudospin state aligned (anti-aligned) with the local internal field ($g_{\rm L} \vec{B}$ in this case).
The electron in the right dot will serve as a reference and is initialized in its local pseudospin ground state $\ket{-R}$.
In terms of the two-electron eigenstates discussed above, the system will thus occupy one of the states $\ket{\pm L,-R}$, which are the $(1, 1)$ ground state and the first- or second-excited state, depending on the relative magnitude of $|g_{\rm L} \vec{B}|$ and $|g_{\rm R} \vec{B}|$.
The task of reading out the left spin is thus equivalent to the task of distinguishing these two states.

Such a classification is usually done by a slow, adiabatic detuning sweep to the ``right'' side of the charge transition, i.e., into the $(0,2)$ charge region, followed by a detection of the final charge state of the right dot:
If one of the two $(1,1)$ states to be distinguished connects adiabatically to the $(0,2)$ state while the other connects to a $(1,1)$ state, then the outcome of a charge sensing measurement of the final state provides unambiguous information about the initial state of the left pseudospin.
This readout mechanism is called PSB readout, as the spin-to-charge conversion is based on the fact that the Pauli principle forbids an aligned spin pair to occupy the single ground-state orbital of the right dot.

Comparing the panels of Fig.~\ref{fig:twoelectronspectraclassf}, we can identify a few different scenarios,
depending on the relative magnitude of $|g_{\rm L} \vec{B}|$ and $|g_{\rm R} \vec{B}|$.
(Spectra such as shown in Fig.~\ref{fig:twoelectronspectraclassf} will look qualitatively the same for $|g_{\rm L} \vec{B}| < |g_{\rm R} \vec{B}|$ and $|g_{\rm L} \vec{B}| > |g_{\rm R} \vec{B}|$, the main difference being the ordering of the levels $\ket{\pm L,\pm R}$; below we investigate both cases while referring to Fig.~\ref{fig:twoelectronspectraclassf}.)
If it happens to be the case that $|g_{\rm L} \vec{B}| < |g_{\rm R} \vec{B}|$ then the two $(1,1)$ states to be distinguished are the ground state and first-excited state.
All four spectra shown in Fig.~\ref{fig:twoelectronspectraclassf} now allow in principle for PSB readout, since in all cases the two lowest $(1,1)$ states connect adiabatically to different charge states in the $(0,2)$ region.
However, for a generic field direction [Fig.~\ref{fig:twoelectronspectraclassf}(a)], spin-to-charge conversion might be more demanding than for the magic field directions: 
First, since in the generic case the excited $(1,1)$ state needs to traverse \emph{two} anticrossings adiabatically, with potentially different coupling parameters, a careful engineering of the detuning pulse shape could be needed.\footnote{If the magnitude of the two coupling parameters is very different, one could also design a pulse shape that results in adiabatic evolution across one of the anticrossings and diabatic evolution across the other, which would again result in good spin-to-charge conversion. This is, however, a rather special situation and making it work would require accurate knowledge about the details of the two $g$-tensors.}
Second, in this case the charge-state readout signal could be obscured due to the fact that the final $(1,1)$ state has a finite spin-singlet component, allowing for relatively fast spin-conserving charge relaxation to the $(0,2)$ ground state.

The situation is rather different when $|g_{\rm L} \vec{B}| > |g_{\rm R} \vec{B}|$.
In that case, the initial $(1,1)$ states $\ket{\pm L,-R}$ to be distinguished are the ground and \emph{second}-excited state (the first-excited state being $\ket{-L,+R}$).
Considering the four spectra shown in Fig.~\ref{fig:twoelectronspectraclassf}, we see that the magic field directions corresponding to positive eigenvalues of $M$ [Fig.~\ref{fig:twoelectronspectraclassf}(b) and \ref{fig:twoelectronspectraclassf}(d)] now create a situation where neither of the two $(1,1)$ states connects adiabatically to the $(0,2)$ state, suggesting that there is no reliable spin-to-charge conversion through adiabatic passage in this case.
[In this case, fast spin-conserving charge relaxation in the $(0,2)$ region could in fact restore the PSB signal.]
The situation for the generic field direction [Fig.~\ref{fig:twoelectronspectraclassf}(a)] is very similar to before:
The two $(1,1)$ states do connect to different charge states, but devising the optimal pulse shape for spin-to-charge conversion could be challenging and fast charge relaxation might obscure the signal.
Finally, if the field points along the magic direction that corresponds to a negative eigenvalue of $M$ [Fig.~\ref{fig:twoelectronspectraclassf}(c)], the lowest two $(1,1)$ states again couple adiabatically to different charge states that have an orthogonal (pseudo)spin structure, thus yielding a proper PSB readout signal.

Combining all the observations made so far, we see that magic field directions corresponding to negative eigenvalues of $M$ are favorable for PSB-based spin readout, independent of the ratio $|g_{\rm L} \vec{B}|/|g_{\rm R} \vec{B}|$ and the spin-conserving charge relaxation rate in the $(0,2)$ region.
Since the relative magnitude of $|g_{\rm L} \vec{B}|$ and $|g_{\rm R} \vec{B}|$ could be hard to control or extract in experiment, one should thus rather search for a magic field direction corresponding to a negative eigenvalue of $M$, e.g., along the lines suggested in Sec.~\ref{sec:singleelectrondegeneracies}.
This will yield good spin-to-charge conversion irrespective of the more detailed properties of $g_{\rm L,R}$.

We emphasize that in this case where the field is along a magic direction corresponding to a negative eigenvalue of $M$, the spectrum is inverted as compared with the ``standard'' level ordering: upon sweeping the detuning, Pauli spin blockade [i.e., no tunneling to (0, 2)] occurs for the first excited (1, 1) state.
With this in mind, we now interpret a recent unexpected experimental observation. 
In Ref.~\onlinecite{Hendrickx2021}, the authors implement spin-to-charge conversion and PSB readout in a DQD, and they observe that ``both antiparallel spin states are blocked, opposite to conventional'' Pauli blockade readout. 
Our interpretation is that the device of Ref.~\onlinecite{Hendrickx2021} is a spin-orbit-coupled DQD whose combined $g$ tensor $M$ has a negative eigenvalue, and this particular observation is made when the magnetic field points approximately to a magic direction corresponding to a negative eigenvalue of $M$. 
In that case, the energy spectrum is qualitatively similar to that shown in Fig.~\ref{fig:twoelectronspectraclassf}(c).

With this in mind, we can now also place the connection between PSB and the ``orientation'' of the spin-orbit coupling in the right context.
In a typical experiment where PSB is used to extract information about the spin-orbit coupling, the DQD is tuned to the $(0,2)$ region and connected to a source and drain contact in such a way that transport through the system depends on the charge cycle $(1,1) \to (0,2) \to (0,1) \to (1,1)$, where still the only accessible $(0,2)$ state is a spin singlet.
Whenever one or more $(1,1)$ states have a vanishing overlap with $\ket{\cal S}$, the system will inevitably enter PSB, resulting in a strongly reduced current.
Measuring the current as a function of the direction of applied magnetic field, a minimum is then usually associated with having the external field aligned with an effective spin-orbit field.
From the reasoning presented above, we see that, in terms of the matrix $M$, one expects a reduced current whenever the magnetic field direction hits one of the magic directions.
These dips in the current are, in fact, equivalent to ``stopping points'' of type (iii) and (iv) as discussed in Ref.~\onlinecite{Qvist2022}, where they were explained, as usual, in terms of the relative orientation of the local Zeeman fields as compared with the direction of a field describing the spin-orbit-induced non-spin-conserving tunneling.
In the present work, we understand these directions in a more ``democratic'' way, as resulting from the basic properties of the combined matrix $M = \tilde g_{\rm R}^{-1} R^{-1} \tilde g_{\rm L}$ that includes all onsite and interdot spin-orbit effects.

\section{Transition patterns among the six classes}\label{sec:transitions}

\begin{figure}
\centering
\includegraphics[width=0.35\textwidth]{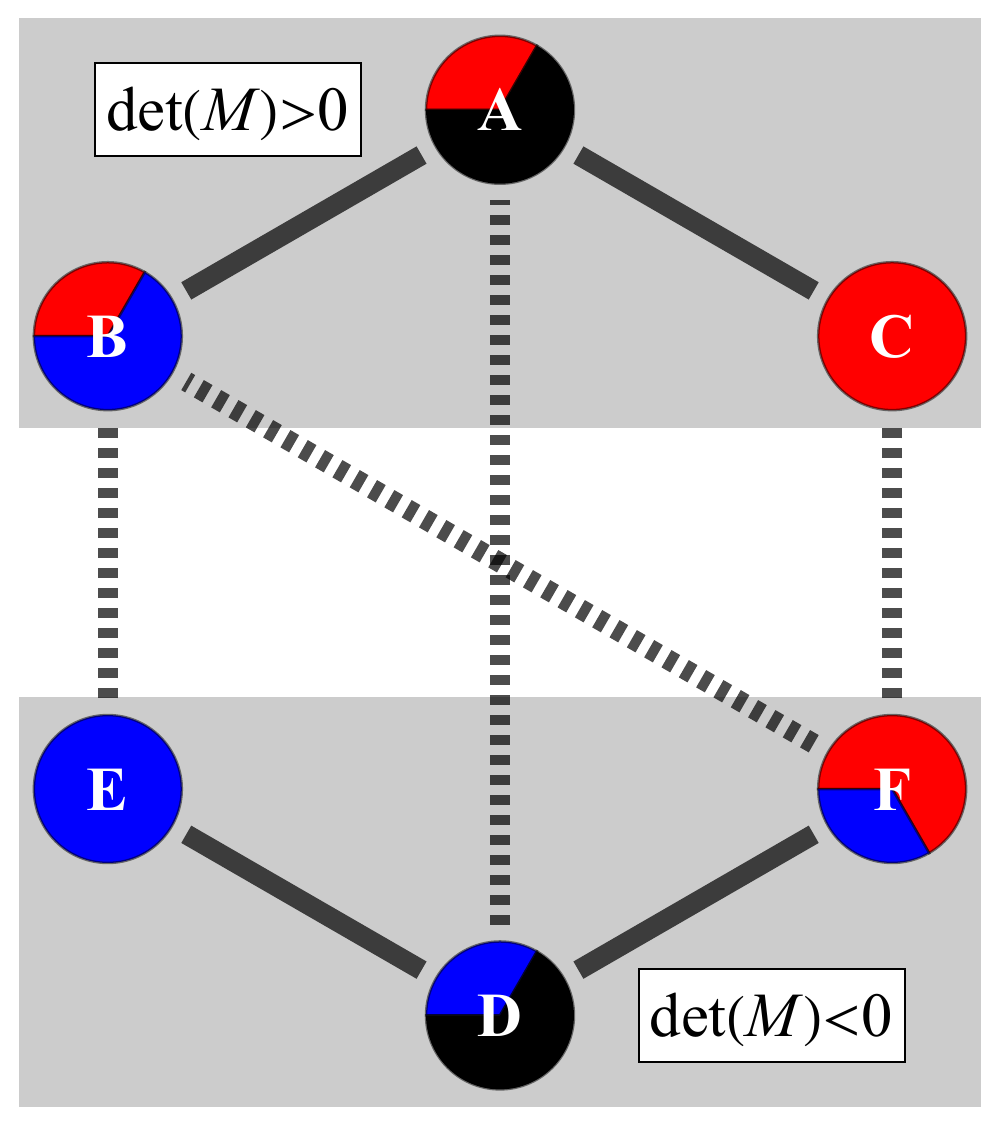}
\caption{Generic transitions among the six classes introduced in Fig.~\ref{fig:eigenexample}. 
Colored circles represent the classes, and their coloring indicates the number of positive/negative/complex eigenvalues of $M$ (red/blue/black).
Solid (dashed) lines indicate generic transitions where the sign of the determinant of $M$ does not change (changes).}
\label{fig:TransitionsPatterns}
\end{figure}

In Sec. \ref{sec:classifications}, we have classified spin-orbit-coupled DQDs into six classes, based on their combined $g$ tensor $M$. 
In an experiment, the two $g$ tensors can be changed in situ, e.g., by changing the gate voltages. 
As a result, the combined $g$ tensor $M$ also changes, and if this change is significant, then $M$ can transition from one class to another. 
Are there any constraints on how $M$ can transition across the classes? 
Yes, there are, as we discuss below. 

We focus on ``generic'' transitions, which require only a single-parameter fine tuning of the $g$ tensors. 
Accordingly, we take into account those cases where one eigenvalue of one of the $g$ tensors goes through zero (without the loss of generality, we assume it is $g_\textrm{L}$), but discard 
more fine-tuned cases, e.g., when \emph{two} eigenvalues of one of the $g$ tensors goes through zero simultaneously, and when one eigenvalue of \emph{each} $g$ tensor goes through zero simultaneously. 

We depict the generic transitions in Fig.~\ref{fig:TransitionsPatterns} as lines connecting the colored circles, where the circles represent the classes.
Solid lines represent transitions where the sign of the determinant of $M$ does not change, whereas dashed lines represent transitions where that sign does change.

In principle, the maximum number of transitions between the six classes could be 15, but we find that only 8 of those transitions are generic, as shown in Fig.~\ref{fig:TransitionsPatterns}.
Instead of a formal proof of this structure, we provide intuitive arguments.

As an example of a generic transition, consider the AB pair of classes, connected by a straight line in Fig.~\ref{fig:TransitionsPatterns}.
It is straightforward to exemplify a process where, by continuously tuning the $g$ tensors,
the two complex eigenvalues shown in Fig.~\ref{fig:eigenexample} (black points) move simultaneously toward the negative real axis, 
collide on the negative real axis, and separate as two different negative eigenvalues.
In fact, tuning the parameter $\alpha$ from 1 to $\pi$ [see inset of Figs.~\ref{fig:eigenexample}(a) and \ref{fig:eigenexample}(b)] does result in such a process.
Furthermore, a small perturbation of such a process still results in a similar change of the 
eigenvalue structure of $M$. 
Hence, the AB transition is generic.

As a counterexample, consider the BC pair of classes, which are not connected in Fig.~\ref{fig:TransitionsPatterns}.
One way to generate this transition is to change the $g$-tensors in such a way that the two negative eigenvalues in Fig.~\ref{fig:eigenexample}(b) (blue points) move to reach zero simultaneously, and then move onto the positive real axis.
Clearly, this requires a higher degree of fine-tuning than a BF transition, where only one of the negative eigenvalues moves across zero.
That is, the BF transition is generic, but the BC transition is not. 
Another way to reach a BC transition is to induce a collision of the two negative eigenvalues to render them a complex pair, and then move them onto the positive real axis. 
This is a BA transition followed by an AC transition.
These arguments illustrate that the BC transition is not generic.

Going beyond such arguments, a formalized derivation of the generic transitions can be given using the codimension counting technique we 
have discussed and used in Sec. III of Ref.~\onlinecite{Frank2020}.
In that language, generic transitions are those that are characterized with a codimension-one eigenvalue pattern of $M$.

\section{Magic loops}
\label{sec:magicloops}

The magic magnetic-field directions we investigated in the previous sections turned out to have many interesting properties, with implications of qualitative importance for the single- and two-electron physics in spin-orbit-coupled DQDs.
As explained in Sec.~\ref{sec:classifications}, these directions, being the eigenvectors of the matrix $M = \tilde g_{\rm R}^{-1} R^{-1} \tilde g_{\rm L}$, are the magnetic-field orientations for which the internal Zeeman fields on the dots are aligned (or anti-aligned).

In the context of PSB, the magic directions result in a proper spin blockade since they make the states $\ket{+L,+R}$ and $\ket{-L,-R}$ truly orthogonal to the pseudospin singlet state.
Aligning the magnetic field along a magic direction is thus expected to fully restore PSB, which is in general lifted in DQDs with strong spin-orbit coupling.
The converse, however, is not true:
A restored spin blockade does not always imply that the external field is pointing along a magic direction.
Indeed, it is known that there is one more internal field configuration, not related to the magic directions, that yields a full spin blockade:
This is the configuration with the two internal fields having equal magnitude, $|g_{\rm L}\vec{B}| = |g_{\rm R}\vec{B}|$.
In this case, the two $(1,1)$ states $\ket{+L,-R}$ and $\ket{-L,+R}$ are degenerate at vanishing interdot tunneling, independent of the relative orientation of the internal fields. 
Both states being tunnel-coupled to the same state $\ket{{\cal S}_{02}}$ will result in one ``bright'' and one ``dark'' state, the latter being fully spin-blocked \cite{Danon2009,Qvist2022}.

A few natural questions arise regarding these equal-Zeeman directions for which $|g_{\rm L}\vec{B}| = |g_{\rm R}\vec{B}|$: 
(i) For a given DQD Hamiltonian, do such equal-Zeeman directions exist?
(ii)
Is their existence determined by the combined $g$ tensor $M$?
(iii) If those equal-Zeeman directions do exist, then how are they arranged on the unit sphere of magnetic-field directions?
(iv) Is there a particular relation between the arrangements of equal-Zeeman directions and the arrangements of the magic directions discussed in previous sections? 
(v) Can we identify any physical consequence of the equal-Zeeman directions, beyond the full PSB discussed above?
We address these questions in what follows.

\subsection{Existence condition of magic loops with equal Zeeman splittings}

The condition of equal Zeeman splittings in the two dots reads:
\begin{equation}\label{eq:loopconditionmagnitude}
    |g_\text L\vec B|=|g_\text R\vec B|.
\end{equation}
This can be rewritten by inserting $g_\text R^{-1}g_\text R$, to obtain
\begin{equation}\label{eq:loop2}
    |g_\text Lg_\text R^{-1}g_\text R\vec B|=|g_\text R\vec B|.
\end{equation}
We introduce the notations
\begin{equation}\label{eq:tildeM}
    M'=g_\text Lg_\text R^{-1},
\end{equation}
and 
\begin{equation}\label{eq:tildeB}
    \vec B'=g_\text R\vec B.
\end{equation}
With this notation, 
Eq.~\eqref{eq:loop2} takes the following simple form:
\begin{equation}\label{eq:loop3}
    |M'\vec B'|=|\vec B'|.
\end{equation}
Rescaling the magnetic-field vector does not change this condition, therefore we rewrite the latter in terms of the unit vector $\vec b' = \vec B'/|{\vec{B}'}|$ characterizing the magnetic field direction.
Then, we obtain
\begin{equation}\label{eq:loop4}
    |M'\vec b'|=1.
\end{equation}

For a given $M'$, is there a unit vector $\vec b'$ that satisfies Eq.~\eqref{eq:loop4}?
This can be answered by analyzing the singular values of $M'$.
We introduce the smallest singular value $\sigma_1$ and the greatest singular value $\sigma_3$ of $M'$:
\begin{equation}\label{eq:singval}
    \sigma_1=\min_{|\vec b'|=1} |M'\vec b'|,\hspace{6mm}\sigma_3=\max_{|\vec b'|=1} |M'\vec b'|.
\end{equation}

According to the defining Eqs.~\eqref{eq:singval}, there is no $\vec b'$ solving Eq.~\eqref{eq:loop4} if 
$\sigma_1>1$ or $\sigma_3<1$.
If, however,  $\sigma_1<1<\sigma_3$, there are unit vectors $\vec b_1'$ and $\vec b_3'$ for which $|M'\vec b_1'|=\sigma_1<1$ and $|M'\vec b_3'|=\sigma_3>1$. This divides the unit sphere of $\vec b'$ into regions where $M'$ contracts or elongates the vectors it acts on. The boundaries between these regions are the unit vectors that satisfy Eq.~\eqref{eq:loop4}. These boundaries appear generally as a pair of loops, related to each other by inversion symmetry. 

\begin{figure}[t]
\centering
\includegraphics[width=0.5\textwidth]{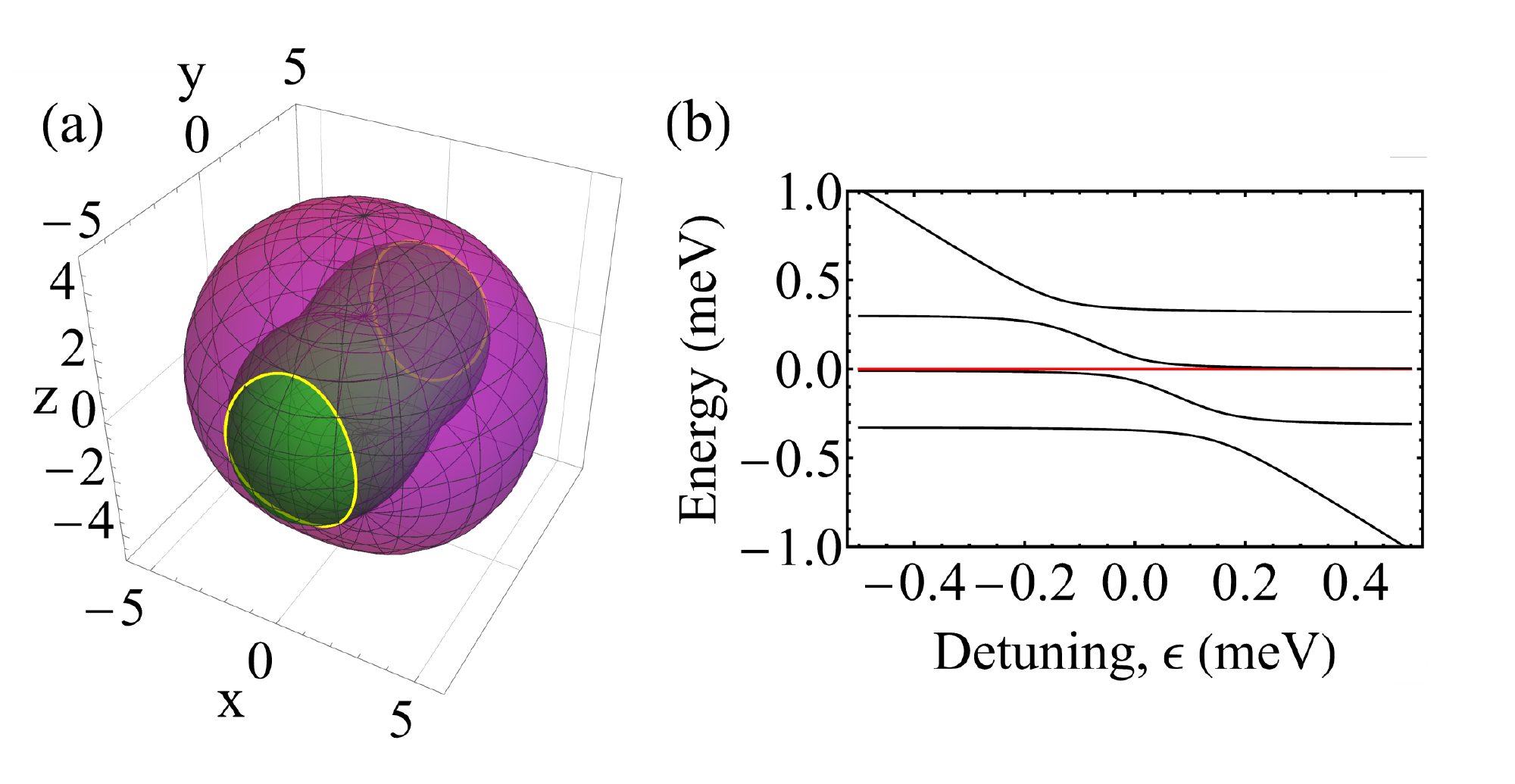}
\caption{Magic loops in a spin-orbit-coupled double quantum dot. 
(a) Dimensionless Zeeman splitting $|g_\text{L}\hat{\boldsymbol{n}}|\hat{\boldsymbol{n}}$ (violet) and $|g_\text{R}\hat{\boldsymbol{n}}|\hat{\boldsymbol{n}}$ (green), of the two dots, where $\hat{\boldsymbol{n}}=\vec{B}/B$ and $B$ is the magnitude of the applied magnetic field. The yellow closed curves are the `magic loops', defined as the intersection curves of the two surfaces. The Zeeman splittings in the two dots are equal for all magnetic field directions denoted by these loops.
The $g$-tensors are $g_\textrm{L} = \diag (6, -4, 5)$ and $g_\textrm{R} = \diag (3, 5, 2)$. 
(b) Detuning-dependent spectrum of a two-electron DQD for a magnetic-field direction that satisfies the magic loop condition in Eq.~\eqref{eq:loopconditionmagnitude}.
We set the tunneling amplitude to ${t}_0 = 0.1$~meV and used
$B = 0.6$~T and $\hat{\boldsymbol{n}} = (1/2, \sqrt{3}/2,0)$. 
A dark state [the flat spectral line at zero energy in panel (b) denoted by red] is formed for all magnetic field directions along the magic loop.
}
\label{fig:magicloop}
\end{figure}

Transforming back, $\vec b = g_{\text R}^{-1}\vec b'$ specifies the magnetic-field directions $\vec b/ |\vec b|$ where Eq.~\eqref{eq:loopconditionmagnitude} is satisfied.
Note that in general, $\vec b$ is not a unit vector and it points to a different direction as $\vec b'$.
However, on the unit sphere of magnetic-field directions, these special directions $\vec{b}/|\vec b|$ also form a pair of loops in an inversion-symmetric configuration. 
We term these loops of equal-Zeeman directions the `magic loops'.
Magic loops are exemplified, for a specific parameter set, as the yellow lines in Fig.~\ref{fig:magicloop}(a), where the violet and green manifolds indicate the relative magnitude of the Zeeman splitting $|g_{\rm L,R} \vec{B}| / |\vec{B}|$ on the left and right dot, respectively, as a function of the direction of $\vec{B}$.
A detuning-dependent spectrum in a two-electron dot, calculated for the magnetic field directed to a point of the magic loop,  is shown in Fig.~\ref{fig:magicloop}(b).
The dark state discussed above is shown in Fig.~\ref{fig:magicloop}(b) as the flat (red) spectral line at zero energy. 

We wish to point out that the definitions of the matrices $M$ and $M'$ are very similar, the only difference being the ordering of $g_\text L$ and $g_{\text R}^{-1}$. The relation between the two matrices is given by the basis transformation \begin{equation}
    M'=g_{\text L}g_{\text R}^{-1}=g_{\text R}g_{\text R}^{-1}g_{\text L}g_{\text R}^{-1}=g_{\text R}Mg_{\text R}^{-1},
\end{equation}
therefore, their eigenvalues are the same. 
Hence, the matrix $M'$ not only determines the existence of magic loops, but it also describes which magic direction class (from A to F) the DQD belongs to, the latter being determined by its eigenvalues. 
The matrix $M$ does not encode both properties, as the singular values of $M$ and $M'$ generally differ.
It is tempting to conclude that, in terms of their information content, $M'$ is superior to $M$, but that is not the case: In fact, the magic directions are determined by $M$ (as the eigenvectors of $M$ corresponding to real eigenvalues), but not determined by $M'$.

So far we defined $M'$ in the gauge of pseudospin-conserving tunneling.
In a generic gauge, the corresponding definition reads: 
\begin{equation}
M'=\tilde{g}_\text L \tilde{g}_\text R^{-1}R^{-1}.
\end{equation}
The eigenvalues and singular values of this $M'$ can be used to perform the magic-direction classification and to determine the existence of the magic loops. 
The rotation $R^{-1}$ is necessary to guarantee the equality of the eigenvalues of $M$ and $M'$. Note that, with this defining equation, $M'$ is a gauge-dependent quantity but its eigenvalues and singular values are not.

\subsection{Stopping points of leakage current in Pauli spin blockade}

Based on the concepts of the (isolated) magic directions and magic loops, we now return to PSB as a dc transport effect as described in the last paragraph of Sec.~\ref{sec:psb}.
Our results imply that [as long as the PSB leakage current is controlled by our Hamiltonian \eqref{eq:twoelectronhamiltonian}] a vanishing leakage current can be caused by the magnetic field being in a magic direction, or being directed to a point of a magic loop. 
One possibility to distinguish between an isolated magic direction and a magic loop is to measure the leakage current in a small region surrounding the original magnetic-field direction.
Another one is to perform detuning-dependent spectroscopy, and identify qualitative features shown either in Fig.~\ref{fig:twoelectronspectraclassf} (magic direction) or in 
Fig.~\ref{fig:magicloop}(b) (magic loop).

\subsection{Dephasing sweet spots}

Finally, we derive another physical property of spin-orbit-coupled DQDs with magic loops, which is practically relevant when the DQD hosts a single electron as a qubit.
We find that if the magic loops are present, and the magnetic field points to a magic-loop direction, then this is a dephasing sweet spot \cite{Piot,Michal} for the qubit at zero detuning, and 
the sweet spot is robust against changing the detuning parameter.

Our derivation relies on the observation that for weak magnetic fields, when both local Zeeman splittings are much smaller than the tunneling amplitude, the splitting between the two lowest eigenstates is described by a detuning-dependent effective or ``averaged'' $g$ tensor, which reads:
\begin{equation}
\label{eq:geff}
    g_\text{eff}(\epsilon) = 
    \frac{1}{2}
    \left[\left(
    1- \frac{\epsilon}{\sqrt{\epsilon^2 + 4 t_0^2}}
    \right) g_L
    +
    \left(
    1+ \frac{\epsilon}{\sqrt{\epsilon^2 + 4 t_0^2}}
    \right) g_R
    \right].
\end{equation}
We assume that, in our case, qubit dephasing is dominated by charge-noise-induced fluctuations of the detuning $\epsilon$.
The defining condition of a dephasing sweet spot is that the fluctuating component of the internal Zeeman field should be perpendicular to the static component.
For our case, this translates to the condition 
\begin{equation}
\label{eq:perpcondition}
\partial_\epsilon g_\text{eff}(\epsilon) \vec B
\perp 
g_\text{eff}(\epsilon) \vec B,
\end{equation}
which is indeed fulfilled if $\epsilon = 0$ and if $\vec B$ is along a magic loop. 
This is proven straightforwardly by performing the derivative of the left-hand side of Eq.~\eqref{eq:perpcondition} using Eq.~\eqref{eq:geff}, evaluating both sides at $\epsilon = 0$, and using the fact that, for three-component real vectors $\vec a$ and $\vec b$ of equal length, $\vec a - \vec b \perp \vec a + \vec b$.

The dephasing sweet spots associated with the magic loops survive a finite static detuning from $\epsilon = 0$.
Our argument for this is as follows: 
We rewrite Eq.~\eqref{eq:perpcondition} as 
\begin{equation}
    f(\epsilon,\theta,\varphi) \equiv
    \left[g_\text{eff}(\epsilon) \hat{\vec n}(\theta,\varphi) \right] \cdot
    \left[
        \partial_\epsilon g_\text{eff}(\epsilon) \hat{\vec n}(\theta,\varphi)
    \right]
    = 0,
\end{equation}
where $\theta$ and $\varphi$ are the polar and azimuthal angles of the magnetic field. 
Consider the detuning value $\epsilon_0 = 0$ where dephasing is reduced for magnetic-field directions along the magic loop, and take a generic point $(\theta_{0},\varphi_{0})$ of the magic loop.
For generic values of the $g$-tensor matrix elements, the derivative $\partial_\varphi f$ does not vanish at $(\epsilon_0,\theta_{0},\varphi_{0})$. 
Therefore, by changing the detuning $\epsilon_0 \mapsto \epsilon_0 + \delta \epsilon$, we can follow the displacement $(\theta_{0},\varphi_{0}) \mapsto (\theta_{0},\varphi_{0}+ \delta \varphi)$ of the corresponding point of the magic loop along the azimuthal direction via
\begin{equation} 
\label{eq:displacement}
    \delta \varphi = - \delta \epsilon
    \frac{ f_\epsilon(0) }{f_\varphi(0)},
\end{equation}
where the new sweet spot is generically slightly away from the magic loop.
Here, we have simplified the notation of the derivatives, e.g., $f_\epsilon(0) \equiv (\partial_\epsilon f)(\epsilon_0, \theta_{0}, \varphi_{0})$.
(Of course, an alternative formulation of this argument is obtained by exchanging the roles of $\theta$ and $\varphi$.)

The mapping of the displacement via Eq.~\eqref{eq:displacement} can be done for all points of the magic loop. 
Hence, we conclude that the dephasing sweet spots identified for zero detuning survive for finite detuning, but the loop formed by these points on the unit sphere of magnetic-field directions is distorted as $\epsilon$ changes. 
Note that, depending on the two $g$-tensors, it may happen that for a finite critical value of $\epsilon$, each loop contracts to a single point.
    
An equivalent argument for the survival of dephasing sweet spots at finite detuning is as follows:
The dephasing sweet spot condition is given by Eq.~\eqref{eq:perpcondition}, valid also for finite detuning.
We now perform the differential on the left-hand side of Eq.~\eqref{eq:perpcondition} using Eq.~\eqref{eq:geff}, and exploit the fact that for real three-component vectors $\vec a$ and $\vec b$, the conditions $\vec a \perp \vec b$ and
$|\vec a + \vec b| = |\vec a - \vec b|$ are equivalent.
This translates Eq.~\eqref{eq:perpcondition} to the following form:
\begin{equation}
\label{eq:displaceddephasingloop}
\left| \left[g_L - G(\epsilon)\right] \vec B\right|
= 
\left| \left[g_R + G(\epsilon)\right] \vec B\right|,
\end{equation}
with
\begin{equation}
    G(\epsilon) = \frac{\epsilon}{2\sqrt{\epsilon^2+4 t_0^2}} 
    \left(
        g_\text{L} - g_\text{R}
    \right).
\end{equation}
Equation \eqref{eq:displaceddephasingloop} has the same structure as Eq.~\eqref{eq:loopconditionmagnitude}, with the only difference being that the matrices in the former are different from the matrices in the latter. 
Furthermore, there is a continuous connection between those matrices, as $G(\epsilon = 0) = 0$.
The singular-value analysis carried out above for Eq.~\eqref{eq:loopconditionmagnitude} can be straightforwardly adopted for Eq.~\eqref{eq:displaceddephasingloop}, yielding $\epsilon$-dependent smallest and greatest singular values $\sigma_1(\epsilon)$ and $\sigma_3(\epsilon)$, both being continuous functions as $G(\epsilon = 0) = 0$. 
This continuity implies that if the magic loops exist, i.e., $\sigma_1(0) < 1 < \sigma_3(0)$, then there is a detuning neighborhood around $\epsilon = 0$ where $\sigma_1(\epsilon) < 1 < \sigma_3(\epsilon)$ holds, and therefore loops of reduced dephasing on the unit sphere of magnetic-field directions do exist.

We remark that the claim of Sec.~\ref{sec:relaxationsweetspot}, i.e., that a magic magnetic-field direction provides a relaxation sweet spot, can be derived using the notion of the averaged $g$ tensor introduced and expressed in Eq.~\eqref{eq:geff}.
Namely, for a magic direction, $g_\text{eff}(\epsilon) \vec B$ is a weighted sum of the two parallel local internal Zeeman fields $g_L \vec B$ and $g_R \vec B$, which implies that the direction of $g_\text{eff}(\epsilon) \vec B$ does not depend on $\epsilon$.
In turn, this implies that a fluctuating internal Zeeman field, caused by a fluctuating detuning, does not have a transversal component to the static internal Zeeman field, which leads to a suppression of qubit relaxation.

\section{Conclusions}
\label{sec:conclusions}

We have proposed a sixfold classification (classes A--F) of spin-orbit-coupled double quantum dots, based on a partitioning of the multidimensional space of their $g$-tensors. 
Our class A includes the often-studied special case of isotropic and identical local $g$-tensors and spin-dependent tunneling due to Rashba- and Dresselhaus-type spin-orbit interaction; the other five classes imply distinct phenomenology to be explored experimentally.
We have argued that the class determines physical characteristics of the double dot, i.e., features in transport, spectroscopy, and coherence measurements, as well as qubit control, shuttling, and readout experiments. 
In particular, we have shown that the spin physics is highly simplified by pseudospin conservation if the external field is pointing to special directions (`magic directions'), where the number of special directions is determined by the class.
We also analyzed the existence and relevance of magic loops in the space of magnetic-field directions, corresponding to equal local Zeeman splittings.
The theoretical understanding our study provides is necessary for the correct interpretation and efficient design of spin-based quantum computing experiments in material systems with strong spin-orbit interaction.

\section*{Acknowledgement}

We thank J.~Asb\'oth, L.~Han, G.~Katsaros, and Y.-M. Niquet for useful discussions. 
This research was supported by the Ministry of Culture and Innovation and the National Research, Development and Innovation Office (NKFIH) within the Quantum Information National Laboratory of Hungary (Grant No. 2022-2.1.1-NL-2022-00004), by NKFIH through the OTKA Grant FK 132146, and by the European Union through the Horizon Europe project IGNITE.
We acknowledge financial support from the ONCHIPS
project funded by the European Union’s Horizon Europe research and innovation programme under Grant Agreement No. 101080022.

\appendix
%\label{appendix}

\bibliography{ref.bib}

\onecolumngrid

\appendix

\section{Gauge transformation}\label{appGaugeTransformation}

In this section, we describe the transformation of the Hamiltonian to a gauge of pseudospin-conserving tunneling, to write the Hamiltonian in the form as in Eq.~\eqref{eq:hamiltoniangaugesingleelectron}. 
We have the freedom to choose the Kramers-pair local basis states on each quantum dot. 
As a consequence of spin-orbit interaction, we have a pseudospin-non-conserving tunneling vector $\tilde{\boldsymbol{t}}$, introduced in Eq.~\eqref{eq:hamiltoniansingleelectron_c}, which describes the rotation of the pseudospin upon the interdot tunneling event. 
We redefine the basis on the right dot to eliminate $\tilde{\boldsymbol{t}}$, which  renders the pseudospin-conserving tunneling energy to $t_0 = \left(\tilde{t_0}^2 + \tilde{\boldsymbol{t}}^2\right)^{1/2}$ (see Eq.~\eqref{eq:hamiltoniansingleelectron_c}).

The tunneling Hamiltonian $H_\text{t}$ in the basis $\ket*{L \tilde \Uparrow}$, 
$\ket*{L\tilde \Downarrow}$, $\ket*{R \tilde \Uparrow}$, and $\ket*{R \tilde \Downarrow}$ has the following matrix form,

\begin{equation}\label{eqappA:TunnelingHamiltonian}
    H_\text{t} = \mqty(0 & 0 & \tilde t_0 - \mi \tilde t_z  & -\mi (\tilde t_x - \mi \tilde t_y) \\ 0 & 0 & -\mi (\tilde t_x + \mi \tilde t_y) & \tilde t_0 + \mi \tilde t_z \\ \tilde t_0 + \mi \tilde t_z & \mi (\tilde t_x - \mi \tilde t_y) & 0 & 0\\ \mi (\tilde t_x + \mi \tilde t_y) & \tilde t_0 - \mi \tilde t_z & 0 & 0).
\end{equation}
The definition of the new basis is the following:
\begin{eqnarray}\label{eqappA:RightBasisDefinition}
\begin{aligned}
&\ket*{R \Uparrow} = \frac{(\tilde t_0 + \mi \tilde t_z)\ket*{R \tilde\Uparrow} + (-\tilde t_y + \mi \tilde t_x)\ket*{R \tilde\Downarrow}}{\tilde{t_0}^2 + \tilde{\boldsymbol{t}}^2} = W\ket*{R \tilde\Uparrow},\\
&\ket*{R \Downarrow} = \frac{(\tilde t_0 - \mi \tilde t_z)\ket*{R \tilde\Downarrow} + (\tilde t_y + \mi \tilde t_x)\ket*{R \tilde\Uparrow}}{\tilde{t_0}^2 + \tilde{\boldsymbol{t}}^2} = W\ket*{R \tilde\Downarrow}.
\end{aligned}
\end{eqnarray}
This basis transformation is a local rotation of the pseudospin around the vector $ \tilde{\mathbf{n}}_\text{so} = \tilde{\mathbf{t}}/|\tilde{\mathbf{t}}|$ with an angle $\theta_\text{so} = 2\arctan\frac{|\tilde{\mathbf{t}}|}{\tilde t_0}$ on the right dot, i.e., $W =e^{i\frac{\theta_\text{so}}{2}\tilde{\mathbf{n}}_\text{so}\tilde{\boldsymbol{\sigma}}_\textrm{R}}$. Note that $W$ has the same form in the new basis as well.

The original basis was formed by a Kramers pair, and the new basis is also formed by a Kramers pair, i.e., $\mathcal{T} \ket*{R \Uparrow} = \ket*{R \Downarrow}$. 
The tunneling Hamiltonian, rewritten in the new basis, reads:
\begin{equation}\label{eqappA:HtunnelNewBasis}
    H_\text{t} = t_0 \ket*{L \tilde \Uparrow}\bra{R \Uparrow} +  t_0 \ket*{L \tilde \Downarrow}\bra{R \Downarrow} + h.c.
\end{equation}

The basis transformation also changes the $g$-tensor on the right dot. The Zeeman Hamiltonian for the right dot in the original basis is proportional to  $\tilde{\boldsymbol{\sigma}}_\text{R} \cdot \tilde g_\text{R}\vec B$. The Pauli matrices in the new basis can be related to the Pauli matrices in the original basis as $\tilde{\boldsymbol{\sigma}}_\textrm{R}
= W^\dag \boldsymbol{\sigma}_\textrm{R} W = R^{-1}\boldsymbol{\sigma}_\textrm{R}$, where $R$ matrix is a three-dimensional rotation with angle $\theta_\text{so}$ around the axis $\tilde{\mathbf{n}}_\text{so}$. The elements of $R$ can be expressed with the tunneling parameters directly:
\begin{equation}
    R_{ij} = \frac{(\tilde{t_0}^2 - \tilde{\boldsymbol{t}}^2)\delta_{ij} + 2 \tilde t_i \tilde t_j - 2 \tilde t_0 \sum_k \varepsilon_{ijk} \tilde t_k}{\tilde{t_0}^2 + \tilde{\boldsymbol{t}}^2}.
\end{equation}
Note that the determinant of $R$ is 1. Thus, the Zeeman term expressed in the new basis is proportional to \mbox{$ R^{-1} \boldsymbol{\sigma}_\textrm{R} \cdot \tilde g_\text{R}\vec B = \boldsymbol{\sigma}_\textrm{R} \cdot g_\text{R}\vec B$}, with the rotated $g$-tensor $g_\text{R} = R\, \tilde{g}_\text{R}$.

\section{Quantum capacitance features along the magic magnetic-field direction}
\label{app:capacitance}

\begin{figure}
    \centering
    \includegraphics[width=0.45\textwidth]{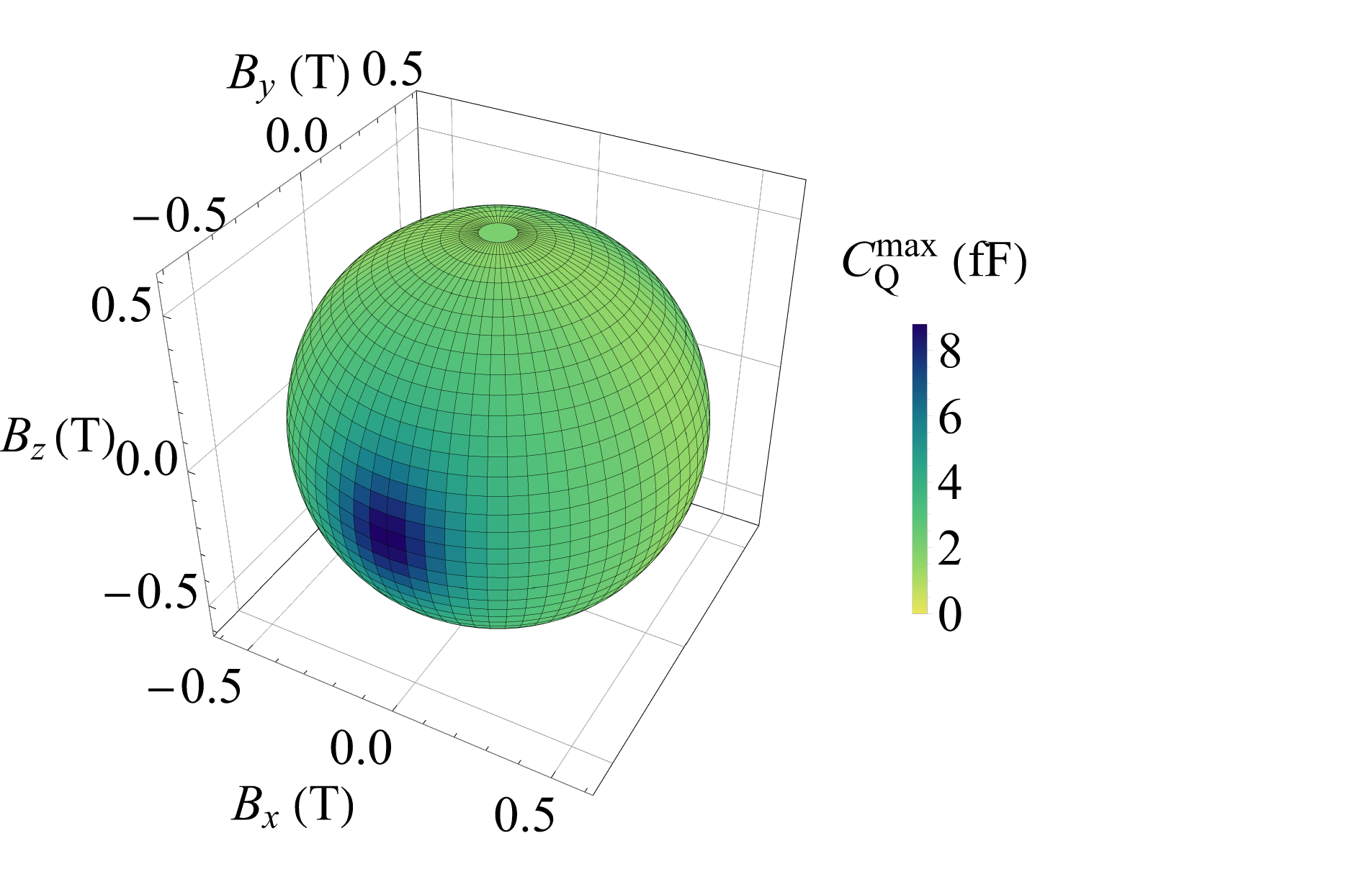}
    \caption{Quantum capacitance peak height as function of magnetic field direction. 
    Quantum capacitance maximum $C_Q^{\text{max}}$ over the detuning axis is plotted, as a function of the magnetic field direction, for a single electron in a double quantum dot at a finite temperature of $50$ mK. 
    The $g$-tensor pair used here belongs to class F, $g_\text{L} = \text{diag}(6,-4,5)$ and $g_\text{R} =\text{diag}(3,5,2)$. The magic direction corresponding to the negative eigenvalue of $M$ is the $y$ axis, and this direction is signalled by the highest value of $C_Q^{\text{max}}$. 
    Further parameters are  $t_0 = 0.02$ meV and $B = 0.6$ T.}  
    \label{fig:appcCq3dsingle}
\end{figure}

Here, we discuss the quantum capacitance features of a spin-orbit-coupled DQD hosting a single electron in thermal equilibrium.
We expect that the measurement of this quantity as a function of magnetic field direction and detuning can reveal a magic direction that belongs to a negative eigenvalue of the combined $g$-tensor $M$. 
This expectation is reinforced by a recent experiment \cite{Han2023}, which demonstrates the quantum capacitance of a two-electron DQD is \emph{suppressed} in the vicinity of a ground-state crossing, as compared with the case when an anticrossing is induced by spin-orbit interaction. 
Interestingly, the simple model we present here predicts an \emph{enhancement} of the quantum capacitance at a crossing point.
We point out mechanisms that are probably relevant to explain this difference, but postpone a detailed analysis for future work. 

We start our analysis with a DQD charge qubit, disregarding spin for simplicity. 
The Hamiltonian in the left-right basis reads
\begin{equation}\label{eqappB:TwoLevelHamiltonian}
    H = \mqty(0 & \Delta/2\\ \Delta/2 & \epsilon),
\end{equation}
where $\epsilon$ is the on-site energy detuning on the two dots and $\Delta$ is the interdot tunneling matrix element.
At finite temperature, the quantum capacitance contribution of the ground state and the excited state are respectively given by
\begin{eqnarray}\label{eqappB:TwoStateCqformulae}
    \begin{aligned}
        &C^{\text{g}}_Q = e^2\frac{\Delta^2/2}{(\epsilon^2 + \Delta^2)^{3/2}}\times\frac{\exp(-\beta\, E_g)}{Z},\\
        &C^{\text{e}}_Q = - e^2\frac{\Delta^2/2}{(\epsilon^2 + \Delta^2)^{3/2}}\times\frac{\exp(-\beta\, E_e)}{Z},
    \end{aligned}
\end{eqnarray}
where $E_g$ and $E_e$ are the two energy eigenvalues, $Z$ is the canonical partition function, $\beta = 1/k_B\, T$, and $e$ is the electron charge.
The total quantum capacitance $C_Q$ is a sum of these two contributions.
(Note that in an experiment, the measurement of this quantum capacitance is also influenced by the lever arms between the gate electrodes and the quantum dots.)
As a function of detuning $\epsilon$, the total quantum capacitance shows a peak at $\epsilon = 0$ with peak width $\propto \Delta$ and peak height expressed as
\begin{equation}\label{eqappB:TwolevelCqmax}
     C^{\text{max}}_Q \equiv 
     \max_{\epsilon} C_Q(\epsilon,\Delta,T)
     =
     \frac{e^2}{2 
     \Delta}\tanh{\frac{\Delta}{2\,k_B\,T}}.
 \end{equation}

Remarkably, this function decreases monotonically as the anticrossing size $\Delta$ increases. 
In the limit of a crossing, i.e., for $\Delta\rightarrow0$, from Eq.~\eqref{eqappB:TwolevelCqmax} we obtain the following expression for the peak height:
\begin{equation}
\label{eqappB:TwoStateCqmaxLimit}
     C^{\text{max}}_Q = \frac{e^2}{4\, k_B\, T}.
\end{equation}
The peak width converges to zero in this $\Delta \to 0$ limit.
Importantly, a quantum capacitance measurement using a radio-frequency probe signal \cite{Mizuta2017,Esterli2019,Vahid2020} might yield a result very different from $C_Q$, especially in the limit of $\Delta \to 0$, as we discuss below. 

Using the approach described above, we computed the thermal-equilibrium quantum capacitance $C_Q(\epsilon,\vec B, T)$ of a single-electron DQD with $g$ tensors, for the example we have analyzed in Fig.~\ref{fig:alignedantialigned}.
For detuning values where the ground state participates in crossings [e.g., Fig.~\ref{fig:alignedantialigned}(d)] or anticrossings [e.g.,  Fig.~\ref{fig:alignedantialigned}(b,c)], the detuning dependence of the thermal quantum capacitance develops a peak with a height essentially described by Eq.~\eqref{eqappB:TwolevelCqmax}. 
Figure \ref{fig:appcCq3dsingle} shows the height of this capacitance peak height $C_Q^{\text{max}} =\max_{\epsilon} C_Q(\epsilon,\vec B,T)$ as the function of the magnetic-field direction, with a fixed magnetic-field strength (see caption for parameters).
This spherical plot exhibits a maximum of $C_Q^{\text{max}}$ along the magic magnetic-field direction corresponding to the negative eigenvalue. 
(Note that our point grid on the spherical surface intentionally avoids the magic direction itself to avoid the $\Delta \to 0$ limit.)

In principle, the pronounced feature observed in Fig.~\ref{fig:appcCq3dsingle} would be useful to identify magic directions corresponding to a negative eigenvalue, using relatively simple thermal-equilibrium capacitance measurements \cite{Mizuta2017,Esterli2019,Han2023}. 
However, in practice, the theoretical model we have applied here probably needs to be refined, the required refinements depending on the hierarchy of frequency and energy scales of the experiment.
A recent experimental result \cite{Han2023} that anticipates this has found a \emph{suppression} of quantum capacitance associated with spectral crossings, in contrast to the theory outlined here which predicts an \emph{enhancement}.

Further scales (beyond anticrossing size $\Delta$ and thermal energy $k_B T$) that probably enter such a refined analysis include the amplitude and frequency of the radiofrequency probe field used to measure the quantum capacitance.
In fact, a radiofrequency probe field of sufficient strength and frequency induces overdrive effects \cite{Vahid2020}, e.g., Landau-Zener transitions between the two levels. 
As a consequence, the measured charge response and the apparent quantum capacitance will deviate from the prediction of the simple thermal equilibrium picture we used above. 
In particular, Landau-Zener transitions are efficient when the radiofrequency probe signal drives the DQD charge through a small anticrossing corresponding to the $\Delta \to 0$ limit discussed above. 
Quantitatively, such a scenario is described by, e.g., the conditions $\epsilon, \Delta \ll A$, $\Delta^2 \lesssim A \omega$, where $A$ is the amplitude of the on-site energy oscillations induced by the probe field, and $\omega$ is the frequency of the probe field. 
The resulting diabatic dynamics is expected to lead to an apparent quenching of the quantum capacitance \cite{Vahid2020}, potentially explaining the findings of Ref.~\cite{Han2023}. 
Furthermore, the strength of charge noise causing detuning jitter, as well as the finite resolution of the detuning mesh used in the experiment, can also play a qualitative role in such a quantum capacitance measurement. 
We postpone the detailed analysis of such refined models to future work.

\section{Calculation of the Chern number of the spectral crossing points}
\label{app:chern}

In the main text, we referred to the Chern number $\mathcal{C}$ associated with the ground-state crossing points of the single-electron spectra. 
There, we considered the three-dimensional parameter space defined by the detuning $\epsilon$ and the magnetic-field angles $\theta = \arctan\left[B_z/\left(B^2_x+B^2_y\right)^{1/2}\right]$ and $\phi = \arctan(B_y/B_x)$, where $B_{x,y,z}$ are the Cartesian components of the magnetic field. 
Here, we outline the procedure to calculate that Chern number $\mathcal{C}$. 
More specifically, this procedure evaluates the Chern number associated with the degeneracy point and the ground-state manifold.

\begin{enumerate}[i.]
    \item  We fix eight points $n_i = n (\epsilon_i, \theta_i, \phi_i)$, $i = 1,\ldots,8$ around the ground-state crossing point to define a cube enclosing the crossing point in the three-dimensional parameter space defined above.
    \item At each vertex $n_i$, we calculate the ground-state wave function $\ket*{\psi_g(n_i)}$ to construct the projector $P_i = \dyad{\psi_g(n_i)}$, which is gauge-invariant.
    \item For each side $s$ of the cube,  we calculate the Berry flux $\mathcal{F_s} = -\text{arg Tr}(P_j P_k P_l P_m)$ through that side, where $j$, $k$, $l$, and $m$ are the integers labeling the vertices of that side. 
    This is in fact the Berry phase associated with the loop of the four ground states~\cite{Asboth2016}. 
    The vertices of a side in the Berry flux formula according to the right-hand rule, such that the surface normal of the side points outward.\\
    \item From the Berry fluxes $\mathcal{F}_s$ $(s=1,\ldots,6)$ of all six sides, we calculate the Chern number via
     \begin{equation} \mathcal{C}=\frac{1}{2\pi}\sum^6_{s=1}\mathcal{F}_s,
     \end{equation}
     {which turns out to be $\pm 1$. A similar scheme to calculate the Chern number is mentioned in the supplementary information of Ref.~\cite{Scherubl2019}.}
\end{enumerate}

\end{document}